\documentclass[%
superscriptaddress,
twocolumn, 
longbibliography,
 amsmath,amssymb,
 aps,
pra,
longbibliography,
floatfix,
nobalancelastpage,
10pt,
]{revtex4-1}

\usepackage{float}
\usepackage{graphicx}
\usepackage{dcolumn}
\usepackage{bm}
\usepackage[mathlines]{lineno}
\usepackage{seqsplit} 
\usepackage{bbm}
\usepackage{color}
\usepackage{comment}
\usepackage{amsthm}
\usepackage{amssymb}
\usepackage{grffile}
\theoremstyle{plain}

\theoremstyle{remark}

\theoremstyle{plain}

\theoremstyle{definition}

\theoremstyle{remark}

\usepackage{seqsplit}

\newcommand{\ket}[1]{|#1 \rangle}
\newcommand{\braket}[2]{\langle #1|#2\rangle}

\usepackage{bm}

\newcommand{\subfigimg}[3][,]{%
	\setbox1=\hbox{\includegraphics[#1]{#3}}
	\leavevmode\rlap{\usebox1}
	\rlap{\hspace*{2pt}\raisebox{\dimexpr\ht1-0.5\baselineskip}{{\bfseries \large\textsf{#2}}}}
	\phantom{\usebox1}
}

\usepackage[colorlinks=true, urlcolor=blue, citecolor=blue,linkcolor=blue,citebordercolor={1 0 0},linkbordercolor={0 0 1}]{hyperref}
\usepackage{listings}
\usepackage{parskip} 
\usepackage{graphicx}

\renewcommand{\eqref}[1]{Eq.~(\ref{#1})} 
\newcommand{\figref}[1]{Fig.~\ref{#1}} 

\usepackage{booktabs}

\definecolor{KH}{rgb}{0.2,0.3,0.9}

\begin{document}

\preprint{APS/123-QED}

\title{Oscillating bound states in non-Markovian photonic lattices}

\author{Kian Hwee Lim}
\email{kianhwee\_lim@u.nus.edu}
\thanks{Equal contribution}
\affiliation{Centre for Quantum Technologies, National University of Singapore, 3 Science Drive 2, Singapore 117543}
\author{Wai-Keong Mok}	
\email{darielmok@caltech.edu}
\thanks{Equal contribution}
\affiliation{Centre for Quantum Technologies, National University of Singapore, 3 Science Drive 2, Singapore 117543}
\affiliation{California Institute of Technology, Pasadena, CA 91125, USA}
\author{Leong-Chuan Kwek}
\affiliation{Centre for Quantum Technologies, National University of Singapore, 3 Science Drive 2, Singapore 117543}
\affiliation{MajuLab, CNRS-UNS-NUS-NTU International Joint Research Unit, UMI 3654, Singapore}
\affiliation{National Institute of Education,
Nanyang Technological University, 1 Nanyang Walk, Singapore 637616}
\affiliation{School of Electrical and Electronic Engineering
Block S2.1, 50 Nanyang Avenue, 
Singapore 639798 }

\begin{abstract}
    It is known that the superposition of two bound states in the continuum
    (BIC) leads to the phenomenon of an oscillating bound state, where
    excitations mediated by the continuum modes oscillate persistently.  We
    perform exact calculations for the oscillating BICs in a 1D photonic
    lattice coupled to a ``giant atom" at multiple points.  Our work is
    significantly distinct from previous proposals of oscillating BICs in
    continuous waveguide systems due to the presence of a finite energy band
    contributing band-edge effects.  In particular, we show that the bound
    states outside the energy band are detrimental to the oscillating BIC
    phenomenon, and can be suppressed by increasing either the number of
    coupling points or the separation between each coupling point.
    Crucially, non-Markovianity is necessary for the existence of oscillating
    BIC, and the oscillation amplitude increases with the characteristic
    delay time of the giant atom interactions.  We also propose a novel
    initialization scheme in the BIC subspace.  Our work be experimentally
    implemented on current photonic waveguide array platforms and opens up
    new prospects in utilizing reservoir engineering for the storage of
    quantum information in photonic lattices.

\end{abstract}

\maketitle
\section{Introduction}
The study of interactions between atoms and photons traces its history all
the way back to the inception of quantum mechanics itself. Since then, we
have acquired a better understanding of atom-photon interactions which underpins the foundation of many quantum technologies such as atomic clocks~\cite{ludlow2015optical} and trapped-ion quantum computers and simulators~\cite{bruzewics2019trapped,duan2010quantum,monroe2021programmable}, which harness the interaction of atoms
with lasers. In many studies of atom-photon interactions, one often makes the dipole approximation~\cite{tannoudji1992atom}, which assumes that the size of the atom is
much smaller than the wavelength of the light. This is especially valid in optical regimes where the length scale of the atom $(\approx 10^{-10} \, \text{m})$ is orders of magnitude smaller than the wavelength of light $(\approx 10^{-7} \, \text{m})$. Under the dipole approximation,
the time taken for the light to pass through a single atom is neglected, thus simplifying the interaction model. In
the field of waveguide quantum electrodynamics (QED)~\cite{liao2016photon,roy2017strongly}, which studies the interactions of atoms with a continuum of bosonic modes in a waveguide, the dipole approximation corresponds to
modelling the atoms as coupled to individual points along the waveguide~\cite{shen_coherent_2005_prl,shen_coherent_2005}.  These
atoms could either be actual atoms~\cite{bajcsy_efficient_2009}, or
artificial atoms like quantum dots~\cite{akimov_generation_2007,alexander_coupling_2016,arcari_near-unity_2014}
and superconducting
qubits~\cite{astafiev_resonance_2010,astafiev_ultimate_2010,abdumalikov_electromagnetically_2010}.
A more complete overview of works done in this vein can be found in
\cite{chang_colloqium_2018,gu_microwave_2017,sheremet_waveguide_2021}. 

However, this paradigm of dipole approximation in waveguide QED was recently broken with the discovery of the so-called ``giant atoms"~\cite{kockum_designing_2014,kockum_quantum_2021}, by coupling each atom to two or more points on the waveguide. This was originally achieved by coupling the superconducting artificial atom (working in the microwave regime) to surface acoustic waves (SAW). Due to the low SAW velocity, for a given frequency the wavelength of sound is no longer assumed to be large compared to the size of the superconducting artificial atom. An alternative method to engineer giant atom coupling is by meandering the transmission line such that the atom interacts with the waveguide at multiple locations~\cite{kockum_decoherence-free_2018,kannan_waveguide_2020}. In these setups, we can
no longer ignore the phase acquired by the light propagating in the 1D
waveguide during the interaction with the giant atoms. Remarkably, by tuning the acquired phase \cite{kockum_designing_2014}, one obtains fascinating phenomena such as prolonged coherence time of a giant atom~\cite{kockum_designing_2014, kannan_waveguide_2020}, decoherence-free interactions between two giant atoms
\cite{kockum_decoherence-free_2018,kannan_waveguide_2020} and the non-exponential
decay of a giant atom \cite{guo_giant_2017,andersson_non-exponential_2019}, which have also been experimentally demonstrated in recent years.

Another novel feature of giant atoms in waveguide QED is the existence of
oscillating bound states in the continuum (BIC), which is a genuine
non-Markovian effect due to the significant time delay for information to
propagate between the various coupling points of a giant
atom~\cite{guo_giant_2017}.  The non-Markovianity manifests as a persistent
oscillation of energy in the waveguide trapped between the coupling
points of a giant atom, which behaves akin to a cavity.  This is in stark
contrast to the irreversible loss of energy from the giant atom to the
waveguide in the Markovian regime.  Thus, these oscillating BICs can
potentially be harnessed to preserve quantum information in a non-Markovian
bath by stabilizing the photonic quantum state, which we will show in this
paper.

In the continuous waveguide case, it is usual to linearize the dispersion relation about the atom's energy, since the coupling to the waveguide is weak~\cite{roy2017strongly}. The waveguide can then be regarded as having a linear dispersion with an infinite bandwidth. Instead of using a continuous waveguide such as a transmission line, we
consider a 1D photonic lattice which acts as the reservoir for the giant
atom. The key difference between the 1D photonic lattice that we consider here, and
the continuous waveguide proposed in \cite{guo_oscillating_2020}, is the
presence of a finite energy band where the band edge becomes significant,
which restricts the allowed BICs. Experimentally, this can be achieved using a photonic waveguide array
where each waveguide is side-coupled to each other via the evanescent field
produced by the photon propagating inside the
waveguides~\cite{jones_coupling_1965,somekh_channel_1973}. This has been
proposed to simulate the non-exponential decay of a photonic giant
atom~\cite{longhi_photonic_2020} which is simultaneously coupled to multiple
lattice sites. We also note that while oscillating BICs have been reported in a discrete lattice system with two giant atoms~\cite{longhi_rabi_2021}, manifesting as an effective Rabi oscillation between the atoms, the novelty of our work lies in only requiring a single giant atom to produce oscillating BICs.  

As we will see, having a finite band gives rise to new conditions for the
oscillating BIC phenomenon, which are distinct from those derived for the
continuous waveguide.  Moreover, we now need to consider the effect of bound
states outside the energy band.  These bound states outside the energy band
are out of the continuum of allowed propagating modes and will henceforth be
called bound states outside the continuum (BOC).  As will be explained in
more detail later, these BOCs are detrimental to quantum information storage
as they are states with an exponentially-decaying wave function around the
coupling points of the giant atom to the 1D photonic lattice.  Hence, even
though it is possible to observe oscillatory behavior in the emitter
excitation probability with BOCs~\cite{ramos_nonMarkovian_2016}, we
distinguish the oscillating BICs which we study here, which allow for perfect
quantum information storage, from oscillations induced by BOCs which do not.
We also show that the oscillating BIC is a consequence of the time-delayed
interactions mediated by the 1D photonic lattice, and that a longer time
delay generally results in a higher amplitude for the BIC which reduces the
information leakage.  A longer time delay also suppresses the unwanted
contributions from the BOCs which hinder the ability of the GA to store and
retrieve quantum information.  This allows us to find the optimal conditions
for oscillating BICs.

This paper is organized as follows: firstly, we introduce the model
Hamiltonian and the theory behind BICs in Sec.~\ref{sec:theory}. Our main theoretical results are presented in Sec.~\ref{sec: oscillating bound states} where we derive the new conditions for oscillating BICs in our system as well as optimal conditions to
minimize the detrimental impact of BOCs.
Thereafter, we present some numerical results in Sec.~\ref{sec:numerics} which support our analytical
calculations. To demonstrate the feasibility of our theoretical results, we propose an experimental implementation of our
work achievable on state-of-the art photonic hardware in Sec.~\ref{sec:experiments}. Finally, we conclude in Sec.~\ref{sec:conclusion} and provide several directions for future research.

\section{Theory}
\label{sec:theory}
\subsection{Model Hamiltonian}
The Hamiltonian for the combined atom-lattice system can be written as $H =
H_a + H_{\text{wg}} + H_{\text{int}}$, given in \eqref{eqn: real space hamiltonian} as (setting $\hbar = 1$)
\begin{subequations}
    \label{eqn: real space hamiltonian}
    \begin{align}
        H_{a} &=  \omega_a a^\dag a + U a^{\dag 2} a^2 \\
        \label{eqn: waveguide chain hamiltonian}
        H_{\text{wg}} &= J\sum_{n=1}^{N-1}(b_n^\dagger b_{n+1} + \text{H.c}) \\
        H_{\text{int}} &= \sum_{j=1}^{M} \rho_{j}(a^\dag b_{n_j} + \text{H.c})
    \end{align}
\end{subequations}
where $a$ is the annihilation operator for the giant atom satisfying the
bosonic commutation relation $[a,a^\dag]=1$, and $b_n$ are the annihilation
operators for the 1D photonic lattice with $[b_n, b_m^\dag] = \delta_{mn}$.
$\omega_a$ is the detuning between the giant atom and the photonic lattice.
The $N$ lattice sites are coupled to each other via a tight-binding
Hamiltonian with interaction strength $J$.  The giant atom is coupled to $M$
arbitrary lattice sites $\{n_1, \ldots, n_M\}$ with strength $\rho_j$,
$j=1,\ldots,M$.  Here, $N$ is chosen to be a large number such that we can
treat the lattice as an infinite 1D chain in both the left and right
directions.  The giant atom has an anharmonicity $U$, which we will take $U
\to \infty$ such that it is equivalent to treating the giant atom as a
two-level system.  An illustration can be found in \figref{fig: hamiltonian
description}.

\begin{figure}
  \centering
  \includegraphics[width=0.45\textwidth, trim={0cm 10cm 1cm 0cm},clip]{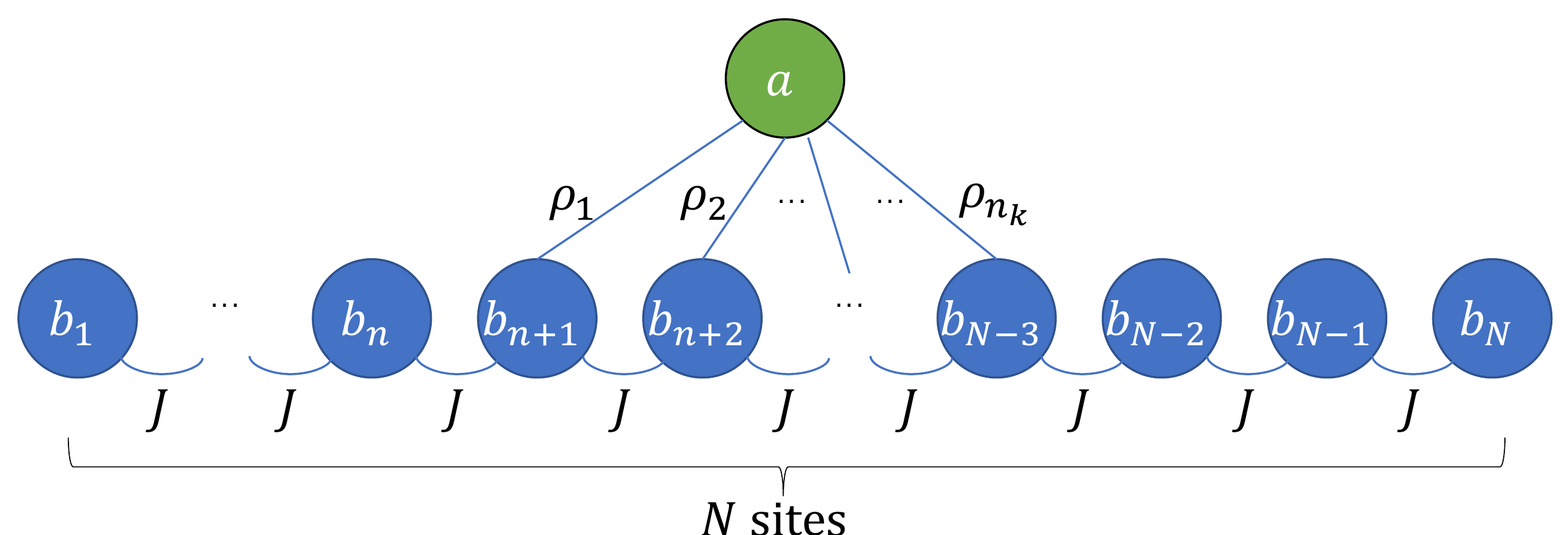}
  \caption[]{An illustration of the Hamiltonian in \eqref{eqn: real space
  hamiltonian}.  Here, we have  $N$ lattice sites arranged in a chain with
  bosonic annihilation operator $b_n$ for each lattice site.  The $N$ lattice
  sites are described by a tight-binding Hamiltonian $H_{\text{wg}}$ with
  coupling strength $J$.  We also have an extra lattice site described by the
  annihilation operator $a$ with Hamiltonian $H_a$, which we shall call the
  ``giant atom'' lattice site, due to its multiple coupling points to the
  1D photonic lattice.  The giant atom lattice site is coupled to the lattice
  sites $n_1, n_2, \dots n_k$ with coupling strengths $\rho_1$, $\rho_2$,
  $\dots$ $\rho_{n_k}$ respectively.  This coupling is described by the
  Hamiltonian $H_{\text{int}}$.  }
  \label{fig: hamiltonian description}
\end{figure}
As is shown in Appendix~\ref{Appendix: k-space hamiltonian derivation}, the
Hamiltonian described by \eqref{eqn: real space hamiltonian} can also be
written in $k$-space in the first Brillouin zone as
\begin{subequations}
\begin{align}
    H_a &= \omega_a a^\dag a \\
    H_{\text{wg}} &= \int_{-\pi}^\pi dk\, \omega(k) c^\dagger(k)c(k) \\
    H_{\text{int}} &= \int_{-\pi}^\pi dk \left\{G(k)a^\dagger c(k) + \text{H.c}\right\}
\end{align}
\end{subequations}
by defining the $k$-space annihilation operators $c(k)$ through the discrete Fourier transform
\begin{equation}
    \label{eqn: k-space annihilation operators}
    c(k) = \frac{1}{\sqrt{2\pi}}\sum_{n=1}^N b_n e^{-ikn}, \quad
    b_n = \frac{1}{\sqrt{2\pi}}\int_{-\pi}^\pi e^{ikn}c(k)\, dk.
\end{equation}
The operators for the lattice $c(k)$ obey the bosonic commutation relations
$\left[c(k),c^\dagger(k^\prime)\right] = \delta(k-k^\prime)$ with the dispersion relation $\omega(k) = 2J \cos(k)$ and the spectral
coupling function $G(k) = \frac{1}{\sqrt{2\pi}} \sum_{j=1}^{M} \rho_j e^{ikn_j}$.  In general, $G(k)$ depends on the specific
geometry of our system, such as the number of coupling points $M$ between the
giant atom and the 1D photonic lattice and also the locations of the
coupling points $n_1, n_2, \dots n_M$. In previous works~\cite{guo_giant_2017,guo_oscillating_2020}, the dispersion relation is linearized and the energy band formed by the waveguide modes is approximated to be infinite such that the band-edge effects become negligible. As we will see later, by confining the allowed energies to be in $[-2J,2J]$, we obtain new conditions for oscillating BICs. Before that, it is helpful to first review the essential physics of BIC in this system.  

\subsection{Bound states in the continuum}
A system is said to have a bound state in the continuum (BIC) if there is an
energy eigenstate with energy $\Omega$, where $\Omega$ lies in the band of
allowed energies of the system.  BICs are theoretically very interesting
because conventionally, we would not expect a bound state to exist within a
continuum of propagating states that would carry the energy of the bound
state away, and yet these BICs truly exist and have been investigated both
theoretically and experimentally~\cite{stillinger_bound_1975,plotnik_experimental_2011,hsu_bound_2016,longhi_bound_2007}. Specifically, for the
setup that we are considering, the system has a BIC with energy $\Omega$ if
$\Omega \in [-2J, 2J]$ which is
defined by the tight-binding dispersion relation $\omega(k) = 2J\cos(k)$.
Furthermore, for a BIC at energy $\Omega$ to exist, either the density of
modes vanishes at $\omega=\Omega$ so that there is no mode in the continuum
for the bound state to decay into, or the coupling to the continuum vanishes
at $\omega=\Omega$.  Lastly, since the BIC is a bound state by definition, we
also require the energy eigenstate at frequency $\Omega$ to have a finite
norm. The above conditions can be stated in more mathematically precise terms~\cite{longhi_bound_2007}.

Defining the density of modes $\rho(\omega) \equiv \frac{\partial k}{\partial
\omega}$, the density of modes vanishing at $\omega=\Omega$ means that we
require $\frac{\partial k}{\partial \omega}\bigr|_{\Omega} = 0$, which is not possible for the tight-binding dispersion relation. Thus, by designing the giant atom coupling, we enforce the
condition for the coupling to the continuum to vanish at $\Omega$
\begin{equation}
    \label{eqn: coupling to continuum vanishes}
    |G(k(\Omega))|^2 = 0 .
\end{equation}
If we restrict ourselves to the one-excitation subspace, a general
time-dependent state of the system can be written as
\begin{equation}
    \label{eqn: one excitation subspace ansatz}
    \ket{\psi(t)} = \psi_a(t) \ket{1_a} + \int_{-\pi}^\pi dk \,
    \psi(k,t)\ket{1_k}
\end{equation}
where $\ket{1_a} = a^\dagger\ket{0}$, $\ket{1_k}=c^\dagger(k)\ket{0}$. By
considering an energy eigenstate $\ket{E}$ also in the one-excitation
subspace, we obtain
\begin{equation}
    \label{eqn: Omega must be in band}
    \Omega - \omega_a = \int_{-\pi}^{\pi}dk 
    \frac{|G(k)|^2}{\Omega - \omega(k)}
\end{equation}
by comparing the coefficients of $\ket{1_a}$ and $\ket{1_k}$ in the energy
eigenvalue equation $H\ket{E}= \Omega\ket{E}$. Hence, the requirement that
we have an energy eigenstate with energy $\Omega$ within the band implies that the solution of \eqref{eqn: Omega must be in band} for
$\Omega$ lies in the range $[-2J,
2J]$.  The preceding calculation also gives us an expression for the
coefficient of $\ket{1_k}$, from which we can deduce that the finite norm requirement of the energy eigenstate is equivalent to $\rho(\omega)|G(k(\omega))|^2$ vanishing in the limit $\omega \to \Omega$ at
least as fast as $\sim(\Omega-\omega)^2$.
The integral in \eqref{eqn: Omega must be in band} can be evaluated by first
evaluating the self-energy $\Sigma(s)$ defined by 
\begin{equation}
    \label{eqn: Sigma(s) equation}
    \Sigma(s) = \int_{-\pi}^\pi dk\,\frac{|G(k)|^2}{is - \omega(k)}.
\end{equation}
from which we get 
\begin{equation}
    \label{eqn: detuning in terms of Sigma(s)}
    \Omega - \omega_a = \text{Re}[\Sigma(s=-i\Omega \pm 0^+)].
\end{equation}
A detailed derivation of the above equations is presented in
Appendix~\ref{Appendix: Omega equation derivation}.

\subsection{Decay dynamics}
In order to probe the decay dynamics of the giant atom into the 1D photonic lattice, we initialize the system with one
excitation in the giant atom and the lattice in the vacuum state.  Mathematically, with reference to \eqref{eqn: one excitation subspace
ansatz}, we have $\psi_a(0)=1$ and $\psi(k,0)=0 \, \forall k\in[-\pi,\pi]$. From the Schr\"odinger equation $i\partial_t \ket{\psi(t)} = H\ket{\psi(t)}$ with these
initial conditions, it can be shown (see Appendix~\ref{Appendix: decay dynamics derivation}) that
$\Sigma(s)$ controls the time-dependent probability amplitude $\psi_a(t)$
through the equation
\begin{equation}
    \label{eqn: decay dynamics}
    \psi_a(t) = \sum_{\text{All residues}} \frac{ie^{st}}{is-\omega_a - \Sigma(s)}.
\end{equation}
From \eqref{eqn: decay dynamics}, we see that poles on the right-hand-side of
the equation with a non-zero real component will lead to a decay in
$\psi_a(t)$.  On the other hand, for the poles on the right-hand-side of
the equation that lie on the imaginary axis, i.e if $s=-i\Omega$, the
exponential factor in the numerator will be $e^{-i\Omega t}$, which is
non-decaying and physically represents a BIC arising from the giant atom
decay. We note that (see Appendix~\ref{Appendix: Omega equation derivation})
when there exists an $\Omega$ that fulfils \eqref{eqn: detuning in terms of
Sigma(s)} as well as $|G(k(\Omega))|^2=0$, then $\Omega$ will be a pole on
the imaginary axis, which means that we will have a BIC at the frequency $\Omega$.

These BICs arising from giant atom decay are very interesting because by
construction they are immune to decay into the 1D lattice, and hence can be
used in a manner analogous to the so-called ``dark states'' for purposes like
storing quantum information
etc~\cite{johnsson_storing_2004,yang_quantum_2004}.  Denoting the BIC
energies as $\Omega_j$, which satisfy both \eqref{eqn: detuning in terms of
Sigma(s)} and \eqref{eqn: coupling to continuum vanishes}, with simple poles
at $s=-i\Omega_j$, we have
\begin{align}
    \psi_a(t) &= \sum_j
    \lim_{s\to-i\Omega_j}\left[\frac{ie^{st}}{is-\omega_a-\Sigma(s)}(s+i\Omega)\right]
    \nonumber \\
   \label{eqn: residue in terms of sigma prime}
    &= \sum_j \frac{e^{-i\Omega_j t}}{1+i\Sigma^\prime(-i\Omega_j)}
\end{align}
where we used L'Hopital's rule and also defined $\Sigma^\prime = \partial_s
\Sigma$ to get to the second line.  Moreover, by noting that
\begin{equation}
    \label{eqn: atom amplitude evolution equation}
    \psi_a(t) = \sum_{E} e^{-iEt}|\braket{1_a}{E}|^2
\end{equation}
we obtain the emitter contribution of each BIC as
\begin{equation}
  \label{eqn: emitter contribution BIC}
  |\phi_a^{(j)}|^2 \equiv |\braket{1_a}{\Omega_j}|^2 = \frac{1}{1+i\Sigma^\prime(-i\Omega_j)}
\end{equation}
The usefulness of each of these BICs
can be quantified by the magnitude of $|\phi_a^{(j)}|^2$, since a large $|\phi_a^{(j)}|^2$
implies that the giant atom has a high probability of
being excited despite the existence of decay channels in the continuum for it to decay into.
\section{Oscillating bound states}
\label{sec: oscillating bound states}
Consider the case of giant atom decay in the one-excitation subspace again.
From \eqref{eqn: decay dynamics}, if there exists two BICs at frequency
$\Omega_\alpha$ and $\Omega_\beta$ that have relative large residues compared to the
other BICs, we have $\psi_a(t) \approx A e^{-i\Omega_\alpha t} + B e^{-i\Omega_\beta t}$
for some complex numbers $A$ and $B$.  This means that the emitter probability
$|\psi_a(t)|^2$ oscillates sinusoidally with frequency $|\Omega_\alpha - \Omega_\beta|/2\pi$. We
can also infer the same fact by looking at \eqref{eqn: atom amplitude
evolution equation}.  In this scenario, we say that our
system exhibits an oscillating BIC.  Interestingly, we will show that these
oscillating BICs inherently require non-Markovianity in the system, resulting
in a bath-induced stabilization of a single-photon quantum state which can be
used both as a photon trapped in a cavity as well as a storage for quantum
information.  Ideally, we would want $|A| = |B|$ so that at some time $t$, we
have $|\psi_a(t)|^2=0$ which means that by turning off the giant atom
couplings to the 1D lattice chain at that time, we can release the stored
photon into the 1D chain.

Let us now calculate the conditions in which the setup shown in
\figref{fig: hamiltonian description} exhibits an oscillating BIC.  Consider
the case where the giant atom has $M$ coupling points equally spaced apart by $n_0$
sites on the photonic lattice. For $N$ lattice sites, let the giant atom be coupled to sites $0$, $n_0$, $2n_0$, $\dots$, $(M-1)n_0$ with a uniform coupling strength $\rho_0$. For this particular setup, we have
\begin{equation}
  G(k) = \frac{\rho_0}{\sqrt{2\pi}} \sum_{j=0}^{M-1} e^{ij kn_0}
\end{equation}
As shown in Appendix~\ref{Appendix: oscillating BIC at n=2 CMI}, it is not
possible for an oscillating BIC to exist when $M=2$, consistent with the
results in a continuous linear waveguide~\cite{guo_oscillating_2020}.  Hence,
we consider the case when $M\geq 3$ for which oscillating BICs exist.
Detailed calculations can be found in Appendix~\ref{Appendix: oscillating BIC calculations}.  We will
summarize some of the key results here.  We first calculate $|G(k)|^2$ to be
\begin{equation}
  \label{eqn: |G(k)|^2 for M legs}
  |G(k)|^2 = \frac{\rho_0^2}{2\pi}\left(M+2\sum_{r=1}^{M-1} (M-r)\cos(kn_0 r)\right)
\end{equation}
which means that when we enforce \eqref{eqn: coupling to continuum vanishes} for the coupling to the continuum to vanish, we have 
\begin{equation}
    \label{eqn: conditions for k for M legs oscillating BS}
    k = \frac{2\pi}{n_0}\left(m\pm \frac{1}{M}\right)
\end{equation}
where $m\in \mathbb{Z}$.  Furthermore, \eqref{eqn: Sigma(s) equation} and
\eqref{eqn: |G(k)|^2 for M legs} together give us
\begin{align}
    \label{eqn: sigma(s) for M legs oscillating BS}
    \Sigma(s) &= \frac{\mp i \rho_0^2}{\sqrt{s^2+4J^2}}\left[M + 
    2\sum_{r=1}^{M-1}(M-r) \alpha^{r n0}\right] \\
    \label{eqn: big term in sigma(s) expression}
    \alpha &= \left(\frac{\mp i\sqrt{s^2+4J^2}+is}{2J}\right)
\end{align}
where we have the negative sign in both $\alpha$ and $\Sigma(s)$ above when
$\text{Re}(s)>0$ and the positive sign when $\text{Re}(s)<0$.  Since the
subsequent results are the same regardless of whether we consider
$\text{Re}(s)>0$ or $\text{Re}(s)<0$, we will restrict ourselves to the
$\text{Re}(s)>0$ case.  Thereafter, from \eqref{eqn: detuning in terms of
Sigma(s)}, \eqref{eqn: sigma(s) for M legs oscillating BS} and \eqref{eqn:
big term in sigma(s) expression} we have
\begin{equation}
  \label{eqn: delta(Omega) for M legs oscillating BS}
  \Omega - \omega_a = \frac{-i \rho_0^2}{\sqrt{4J^2-\Omega^2}} \left[M +
  2\sum_{r=1}^{M-1}(M-r) (e^{i n_0\theta})^r\right]
\end{equation}
where $\theta = \arctan(-\sqrt{4J^2-\Omega^2}/\Omega)$.  Hence, to obtain an
oscillating BIC, we need to solve \eqref{eqn: delta(Omega) for M legs
oscillating BS} together with \eqref{eqn: conditions for k for M legs
oscillating BS} to obtain two eigenenergies $\Omega_1$ and $\Omega_2$ such
that the coupling to the continuum vanishes at these two energies.

At this juncture, we consider the case where $\omega_a=0$, such that the
giant atom energy is positioned at the band center.  This is done so that the
BIC energies $\Omega_1, \Omega_2$ are symmetric about the band center, i.e
$\Omega_1 = - \Omega_2$, which is necessary to obtain perfectly sinusoidal
oscillations of the giant atom emitter probability. From \eqref{eqn: conditions for k for M legs oscillating BS}, we see that $n_0$ being an odd number will not give us BIC energies that are symmetric about the band centre. Hence, $n_0$ is
restricted to be an even number which give us two possibilities, $n_0 =
2(2l)$ or $n_0 = 2(2l+1)$ where $l\in \mathbb{Z}^+$.  Since the giant atom
energy is positioned at the band center, we would expect the BIC energies
that are closer to the band center to correspond to states that have a larger
emitter probability.  Using that criteria, we find that the optimal condition
for oscillating BIC occurs for the $n_0 = 4l$ case (see Appendix~\ref{Appendix: oscillating BIC calculations}) from which we get the BIC energies $\pm \Omega_{\text{BIC}}$ given
by
\begin{equation}
  \label{eqn: BIC energies M legs n0=4l}
  \Omega_{\text{BIC}} = 2J \sin\left(\frac{2\pi}{Mn_0}\right).
\end{equation}
We can then obtain the corresponding $\rho_0$ for $\pm \Omega_{\text{BIC}}$
by substituting \eqref{eqn: BIC energies M legs n0=4l} into \eqref{eqn:
delta(Omega) for M legs oscillating BS} together with $\omega_a=0$ to obtain
\begin{equation}
  \label{eqn: BIC rho0 M legs n0=4l}
  \left(\frac{\rho_0}{J}\right)^2 = \frac{2}{M}\tan\left(\frac{\pi}{M}\right)\sin\left(\frac{4\pi}{M n_0}\right)
\end{equation}
resulting in an oscillating BIC at the frequency $\Omega_\text{BIC}/ \pi$.
Now we can obtain the emitter probabilities of each BIC by substituting
\eqref{eqn: BIC energies M legs n0=4l} and \eqref{eqn: BIC rho0 M legs n0=4l}
into \eqref{eqn: emitter contribution BIC} to obtain
\begin{widetext}
\begin{equation}
  \label{eqn: BIC emitter probs M legs n0=4l}
  |\phi_a^{(\text{BIC})}|^2 \equiv \braket{1_a}{\pm \Omega_{\text{BIC}}}|^2 =
  \frac{i e^{-\frac{4 i \pi }{M}} \left(1+e^{\frac{2 i \pi }{M}}\right) \left(-1+\left(i e^{\frac{2 i \pi }{M
  n_0}}\right)^{n_0}\right)^3 \csc ^2\left(\frac{\pi }{M}\right) \cos ^2\left(\frac{2 \pi }{M n_0}\right)}{4 \left(2 n_0
  \sin \left(\frac{4 \pi }{M n_0}\right)+\sin \left(\frac{2 \pi  (n_0-2)}{M n_0}\right)+\sin \left(\frac{2 \pi 
  (n_0+2)}{M n_0}\right)\right)}
\end{equation}
\end{widetext}
From the emitter probabilities obtained above, using \eqref{eqn: decay
dynamics} and \eqref{eqn: residue in terms of sigma prime}, we see that in
the long time limit after all the non-BIC states have propagated away from
the giant atom to the left and right ends of the lattice chain, we would
expect oscillations in the emitter probability of the giant atom with
amplitude $(2|\phi_a^{(\text{BIC})}|^2)^2$.  From \eqref{eqn: BIC emitter
probs M legs n0=4l}, we note that for all values of $n_0 = 4l, l\in
\mathbb{Z^+}$, $|\phi_a^{(\text{BIC})}|^2$ increases monotonically with $M$,
eventually saturating at the limiting value $|\phi_a^{(\text{BIC})}|^2 = 
1/3$.  Consequently, for any value of $n_0$, having a larger number of
coupling points $M$ leads to higher-amplitude oscillating BICs.
It is also helpful to use \eqref{eqn: BIC emitter probs M legs n0=4l} to
compute the asymptotic behavior of $|\phi_a^{(\text{BIC})}|^2$ as $n_0 \to
\infty$, which we can write as
\begin{align}
  |\phi_a^{(\text{BIC})}|^2 
  \label{eqn: asymptotic emitter prob as n0=4l, n0 tends to infty}
  &= \frac{1}{1+\frac{4\pi}{M}\csc\left(\frac{2\pi}{M}\right)} + \frac{A}{n_0^2} + \mathcal{O}\left(\frac{1}{n_0^4}\right) \\
  \text{where }A&=  \frac{4\pi^2\sin\left(\frac{2 \pi }{M}\right) \left(3 M
  \sin \left(\frac{2 \pi }{M}\right)-4\pi\right)}{3 M \left(M \sin
  \left(\frac{2 \pi }{M}\right)+4 \pi \right)^2}. \nonumber
\end{align}

By defining $\tau=(Mn_0)/\nu_g$ as the time taken for the photon to propagate
between the first coupling point and the $M$th coupling point (i.e, the size
of the giant atom), where $\nu_g=2J$ is the group velocity at the band
centre, and $\Gamma^{-1} = (M^2 \rho_0^2/J)^{-1}$ as the characteristic
timescale for the giant atom decay, we can quantify the amount of
non-Markovianity in our system through the quantity $\tau/\Gamma^{-1}$ which
can be written as
\begin{align}
    \label{eqn: nonMarkovianity measure}
    \frac{\tau}{\Gamma^{-1}} &= M^2 n_0 \sin\left(\frac{4\pi}{M n_0}\right)\tan\left(\frac{\pi}{M}\right) \\
    \label{eqn: nonMarkovianity asymptotic measure}
    &= 4M\pi\tan \left(\frac{\pi }{M}\right)-\frac{32 \left(\pi ^3 \tan
    \left(\frac{\pi }{M}\right)\right)}{3 M
    }\frac{1}{n_0^2}+\mathcal{O}\left(\frac{1}{n_0^4}\right)
\end{align}
where to get from the first line to the second line, we computed the
asymptotic behavior as $n_0 \to \infty$.  From \eqref{eqn: nonMarkovianity
asymptotic measure}, we see that at large $n_0$ the non-Markovianity in our
system, which has the same $1/n_0^2$ scaling as the expressions for the BIC
emitter probabilities in \eqref{eqn: asymptotic emitter prob as n0=4l, n0
tends to infty}.  This allows us to conclude that a stronger non-Markovianity
in our system arising from the time delay for information to propagate
between the giant atom coupling points results in better oscillating BICs,
though the amount of non-Markovianity quantified by $\tau/\Gamma^{-1}$
eventually reaches a plateau.  The presence of the plateau means that even
though $\tau$ increases as $n_0$ increases, which leads to a greater
non-Markovianity in the system, this effect is quickly balanced by an
increase in the giant atom lifetime $\Gamma^{-1}$ which is a result of a
decreased coupling strength $\rho_0$.  In practice, $n_0$ should of course
not be too large since the oscillation period scales as $\sim n_0$ which
might lead to more decoherence.  Fortunately, the fast convergence
$\mathcal{O}(1/n_0^2)$ of the emitter probabilities means that a moderate
$n_0$ is already sufficient to observe good oscillating BICs.

Finally, we note that for all values of
$n_0 = 4l, l\in \mathbb{Z}^+$, as $M\to \infty$, $\tau/\Gamma^{-1}$ monotonically
decreases to a limiting value of $4\pi^2$. This implies that our system with an
oscillating BIC is inherently non-Markovian in nature, since there is a
non-negligible lower bound to $\tau/\Gamma^{-1}$.

\subsection{Role of imperfections: Bound states outside the continuum} 
Bound states outside the continuum (BOCs) are energy eigenstates of the
Hamiltonian that have energy out of the range $[-2J,2J]$.  For these states,
the wave number $k$ is complex~\cite{munro_optical_2017}, which means that
these states are unable to propagate in the 1D lattice chain and hence they
have a significant  probability amplitude in $\ket{1_a}$, with an exponentially-decaying wavefunction around the coupling points of the giant atom. These states are imperfections to our
oscillating BIC for two reasons.  Firstly, for an oscillating BIC produced by
giant atom decay, we want the emitter probability to be high for the two BICs
involved at $\pm \Omega_\text{BIC}$, and low for all the other energy eigenstates.
Yet these BOCs have a large atomic component and hence they act as
imperfections to our sinusoidal oscillation as per \eqref{eqn: decay
dynamics}.  Secondly, these states leak energy outside the giant-atom
coupling points due to the exponential decay of the photon amplitude from the
coupling points.

To characterize the effect of BOCs on oscillating BIC produced by giant
atom decay, we first use \eqref{eqn: BIC rho0 M legs n0=4l} to obtain
$\rho_0$ corresponding to the oscillating BIC condition.  Then, we solve for the BOC energies
$\Omega_{\text{BOC}}$ where $|\Omega_{\text{BOC}}|>2J$ in \eqref{eqn: Omega must be in band} with
$\omega_a=0$.  In the limit of $n_0 \to \infty$, the two BOC energies can be
found as
\begin{equation}
    \label{eqn: asymptotic BOC energies}
    \Omega_{\text{BOC}} \approx \pm 2J \left(1 + \frac{2\pi^2 \tan(\pi/M)^2}{M^2n_0^2} \right)
\end{equation}
with the emitter probability 
\begin{equation}
  \label{eqn: asymptotic BOC probability}
  \left|\phi_a^{(\text{BOC})}\right|^2 \sim \frac{4\pi^2 \tan(\pi/M)^2}{M^2}\frac{1}{n_0^2}.
\end{equation}
This means that for a given value of $M$, at large $n_0$, the contributions
from the BOCs to the oscillations are suppressed by a factor of $1/n_0^2$.
By comparing \eqref{eqn: asymptotic BOC probability} and \eqref{eqn:
nonMarkovianity asymptotic measure}, we see that a larger non-Markovianity in
our system characterized by a larger $\tau/\Gamma^{-1}$ leads to reduced
imperfections from the BOC.  We also note that having a larger number of
coupling points $M$ leads to a diminished effect of the BOCs on our
oscillating bound states, which can be explained by how a larger value of $M$
leads to a smaller coupling $\rho_0$ between the giant atom and the lattice
chain.

\subsection{Initialization in the BIC subspace}
Up till now, we have considered the case of giant atom decay into the 1D
lattice chain.  If instead we are given the ability to initialize the state
of the lattice sites in the chain, which is possible for some experimental
platforms such as a side-coupled waveguide array through pulse shaping
techniques, we can eliminate the effects of the BOCs even at low $n_0$ and
also obtain perfect storage of quantum information within the legs of the
giant atom.  This is especially important for small values of $M$ like $M=3$,
since from \eqref{eqn: asymptotic BOC probability} we see that the emitter probability of the BOC states increases as $M$ decreases.
Hence, we shall temporarily restrict ourselves to $M=3$ here, though it
should be clear that the method below generalizes for any positive integer
values of $M$.

We first write the states corresponding to the BICs at $\pm
\Omega_{\text{BIC}}$ as
\begin{equation}
    \ket{\pm} = \phi_{a,\pm} \ket{1_a} + \sum_{n} \phi_{n,\pm} \ket{n}.
\end{equation}
Without loss of generality, we can set $\phi_{a,\pm} =
|\phi_a^{(\text{BIC})}|$ since
eigenstates are defined up to a global phase.  Here $\phi_{n,\pm}$ are the photon amplitudes in real
space, which for $M=3$ we can calculate $\phi_{n,\pm}$ to be given by
\begin{equation}
  \label{eqn: M=3 bic probability dist}
    \phi_{n,+} = C \times  \begin{cases} \begin{aligned} &(-1)^n \exp{\left( i\frac{2\pi n}{3 n_0}\right)} \\&- \exp{\left( -i\frac{2\pi n}{3 n_0}\right)}, 0\leq n \leq n_0 \\ &(-1)^{n+1} \exp{\left( i\frac{2\pi}{3}\left(\frac{n}{n_0}-2\right)\right)} \\&+ \exp{\left( -i\frac{2\pi}{3}\left(\frac{n}{n_0}-2\right)\right)}, n_0\leq n \leq 2n_0 \\& 0, \quad \text{else} \end{aligned} \end{cases} 
\end{equation}
where
\begin{equation}
    C = \frac{-i^{n+1} \phi_a \rho_0}{2J \cos\left( \frac{2\pi}{3 n_0} \right)}
\end{equation}
and $\phi_{n,-} = (-1)^{n+1} \phi_{n,+}^*$.  The above calculation also means that the state 
\begin{align}
    \label{eqn: BIC subspace state p}
    \ket{p} &\equiv \frac{1}{\sqrt{2}}(\ket{+}-\ket{-}) \nonumber \\
    &= \frac{1}{\sqrt{2}} \sum_n \left(\phi_{n,+}-\phi_{n,-}\right)\ket{n}
\end{align}
is a state with no probability amplitude in $\ket{1_a}$ and with photon
amplitudes in real space only within the $M$ coupling points of the giant
atom.  Thus if we initialize the state in $\ket{p}$, then there will be zero excitation leakage outside of the giant atom and the lattice
sites within the $M$ coupling points.  Furthermore, since the states
$\ket{\pm}$ are orthogonal to the BOCs, the imperfections in the oscillations
due to the BOCs are eliminated by construction.  Lastly, we note that for $M=3$
the photon amplitudes are real and only differ in phase by $0$ or $\pi$, which makes it more feasible for practical implementation.

\section{Numerical results}
\label{sec:numerics}
\begin{figure*}
  \centering
  \subfigimg[width=0.48\textwidth, trim={0cm 0cm 1.5cm 0cm},clip]{(a)}{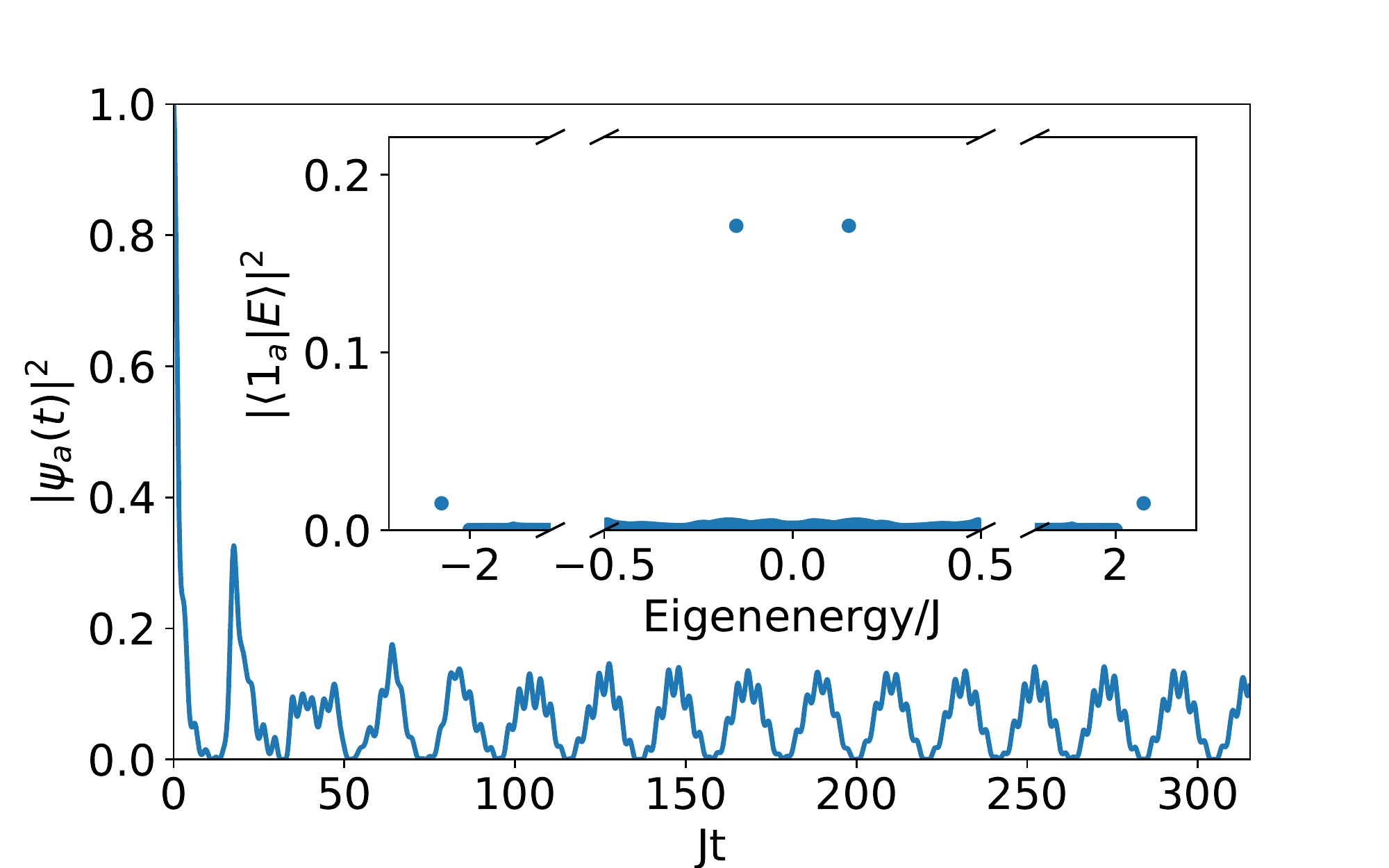}\hfill
  \subfigimg[width=0.48\textwidth, trim={0cm 0cm 1.2cm 0cm},clip]{(b)}{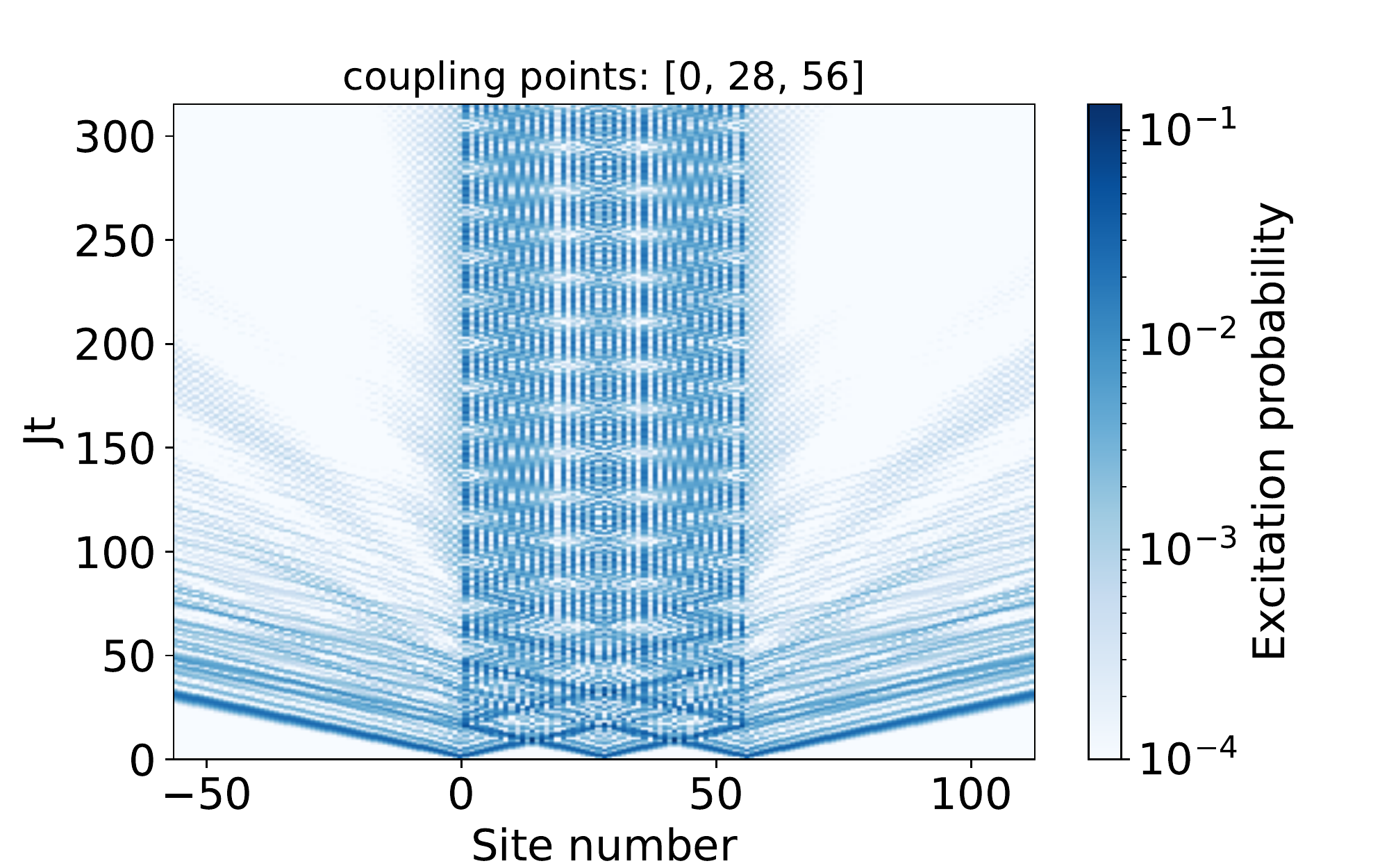}\hfill \\
  \subfigimg[width=0.48\textwidth, trim={0cm 0cm 1.5cm 0cm},clip]{(c)}{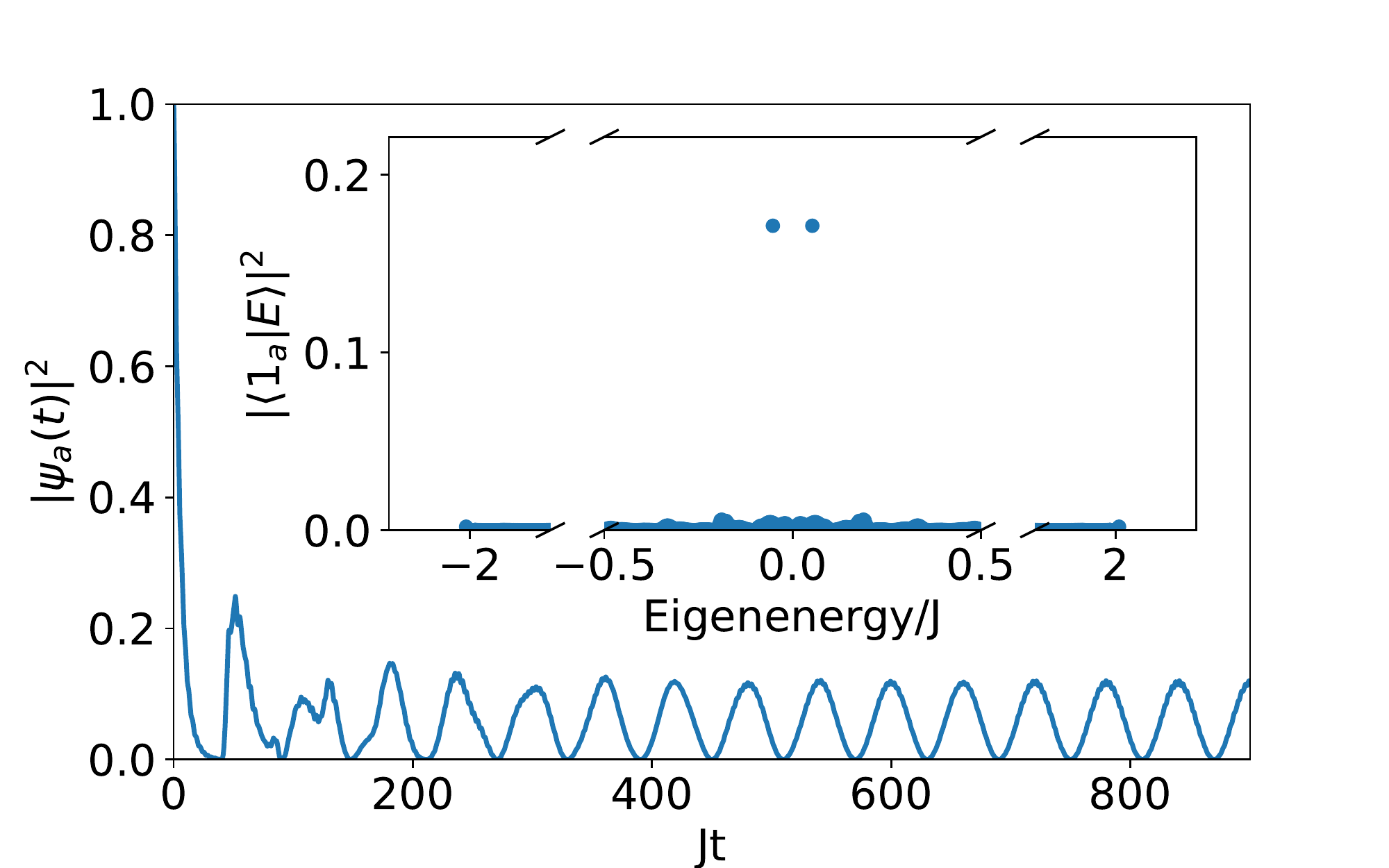}\hfill
  \subfigimg[width=0.48\textwidth, trim={0cm 0cm 1.2cm 0cm},clip]{(d)}{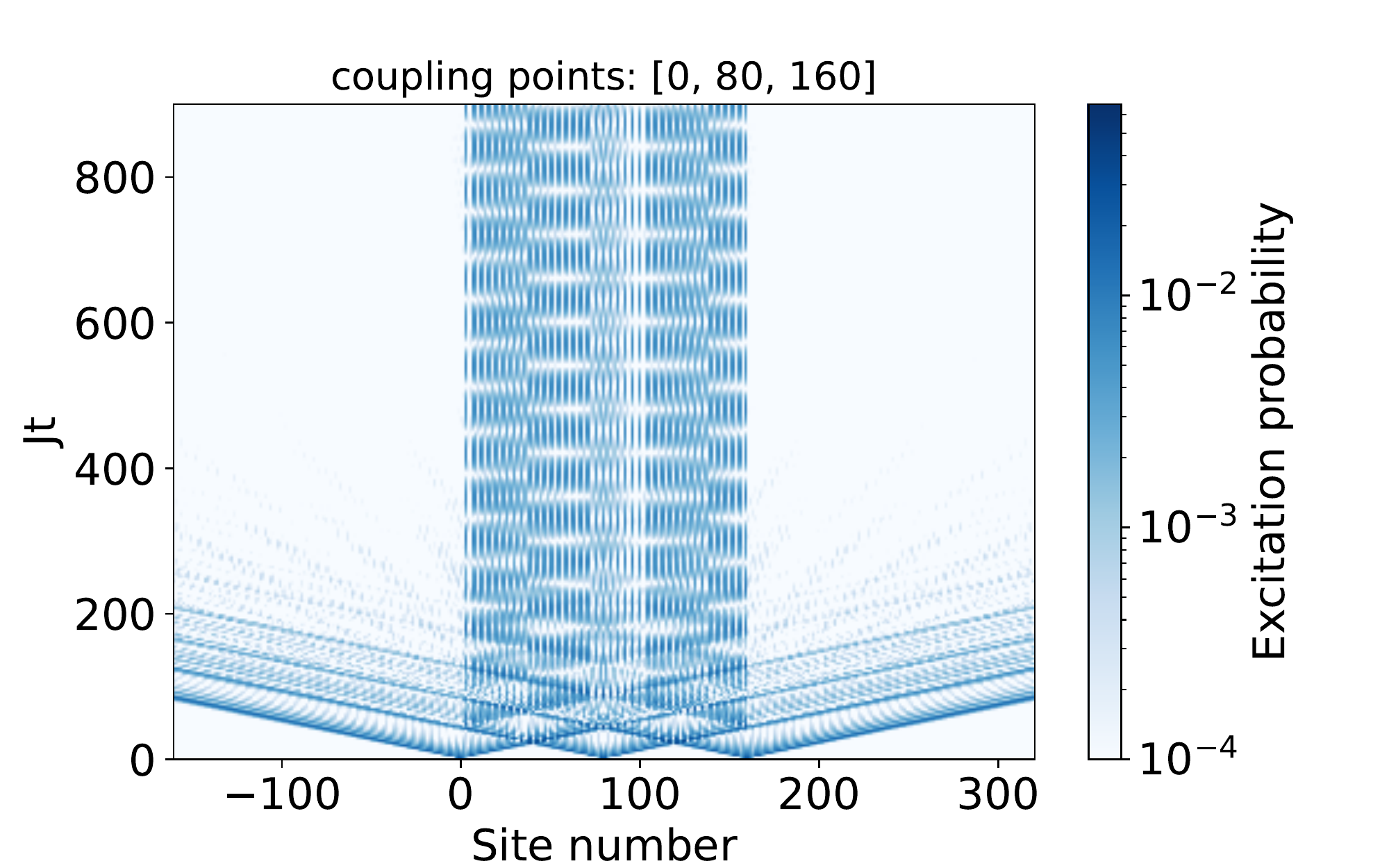}\hfill 
  \caption[]{Simulation results for giant atom decay into the 1D photonic
  lattice for the case of $M=3$ coupling points.  The coupling strength
  $\rho_0$ between the giant atom and the 1D photonic lattice can be
  calculated from \eqref{eqn: BIC rho0 M legs n0=4l}.  In (a) and (c) we have
  the excitation probability against time for the giant atom lattice site for
  $n_0=28$ and $n_0=80$ respectively.  In the insets, we plot the emitter
  probability for each of the eigenstates of $H$.  As can be seen in the
  insets, there are two eigenstates with energies in the continuum $[-2J,2J]$
  that are symmetrical about $0$ and have a emitter probability.  In (b) and
  (d) we have the excitation probability against time for some of the lattice
  sites in the 1D photonic lattice for $n_0 = 28$ and $n_0 = 80$
  respectively.  Initially we see some transient behavior as the non-bound
  states decay into the 1D lattice chain and are propagated away to the left
  and right ends of the 1D photonic lattice.  Thereafter, we see the photon
  being trapped in between the $M=3$ coupling points with non-zero
  probability and oscillating between these three points with time.  By
  comparing (a) and (c), we see that the $n_0=28$ case has oscillations that
  aren't perfectly sinusoidal, due to the effect of the relatively large BOC
  emitter probabilities, as can be seen in the inset of (a).  As was
  previously explained and is seen here in (b), the presence of these BOCs with
  large emitter probabilities for the $n_0=28$ case also lead to a leakage of
  the photon excitation probability beyond the coupling points of the giant
  atom.  We can also see that this imperfection is absent for the $n_0 = 80$
  case.}
  \label{fig: M=3GAdecaySimulationResults}
\end{figure*}

\begin{figure*}
  \centering
  \subfigimg[width=0.48\textwidth, trim={0cm 0cm 1cm 0cm},clip]{(a)}{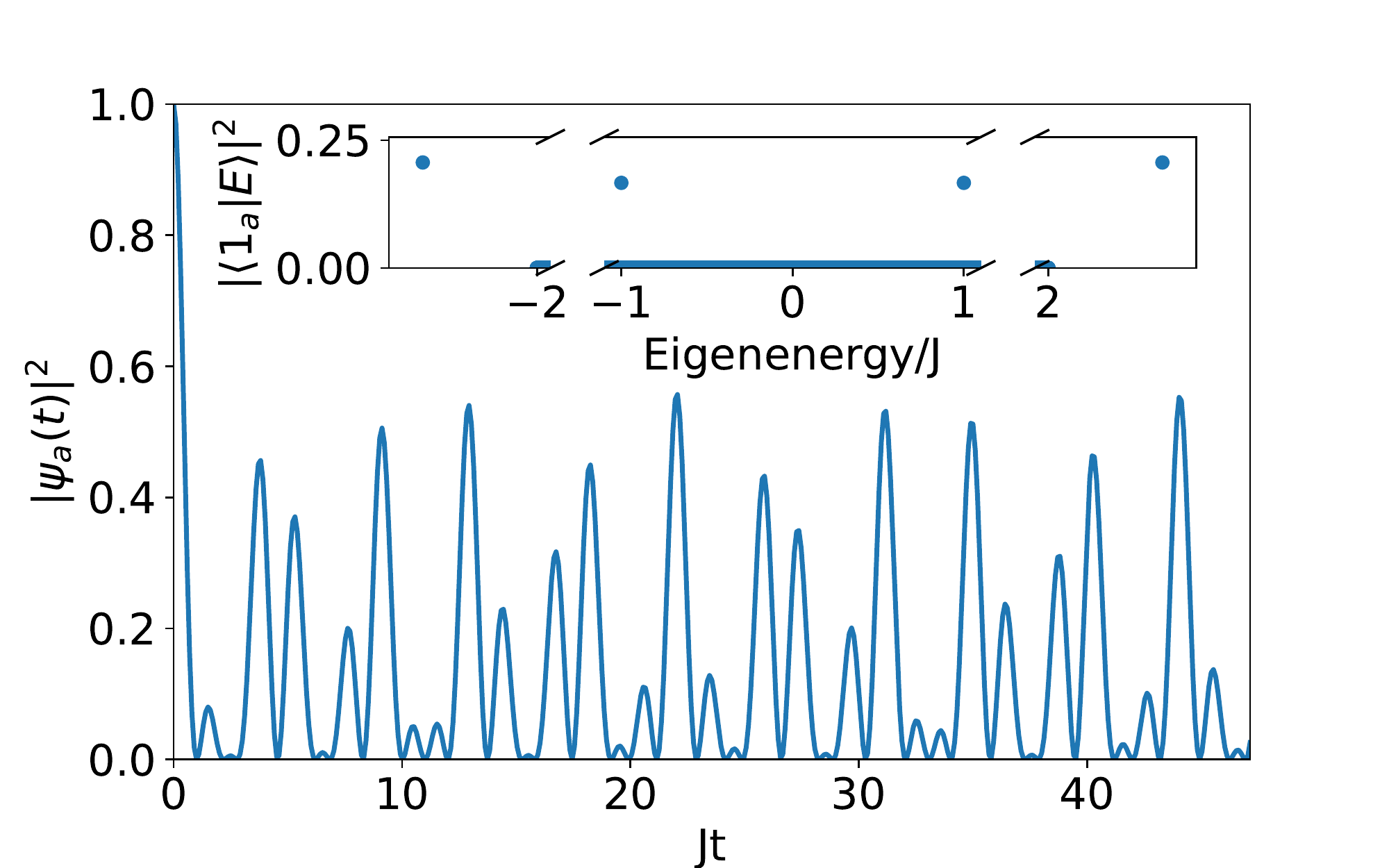}\hfill
  \subfigimg[width=0.48\textwidth, trim={0cm 0cm 1cm 0cm},clip]{(b)}{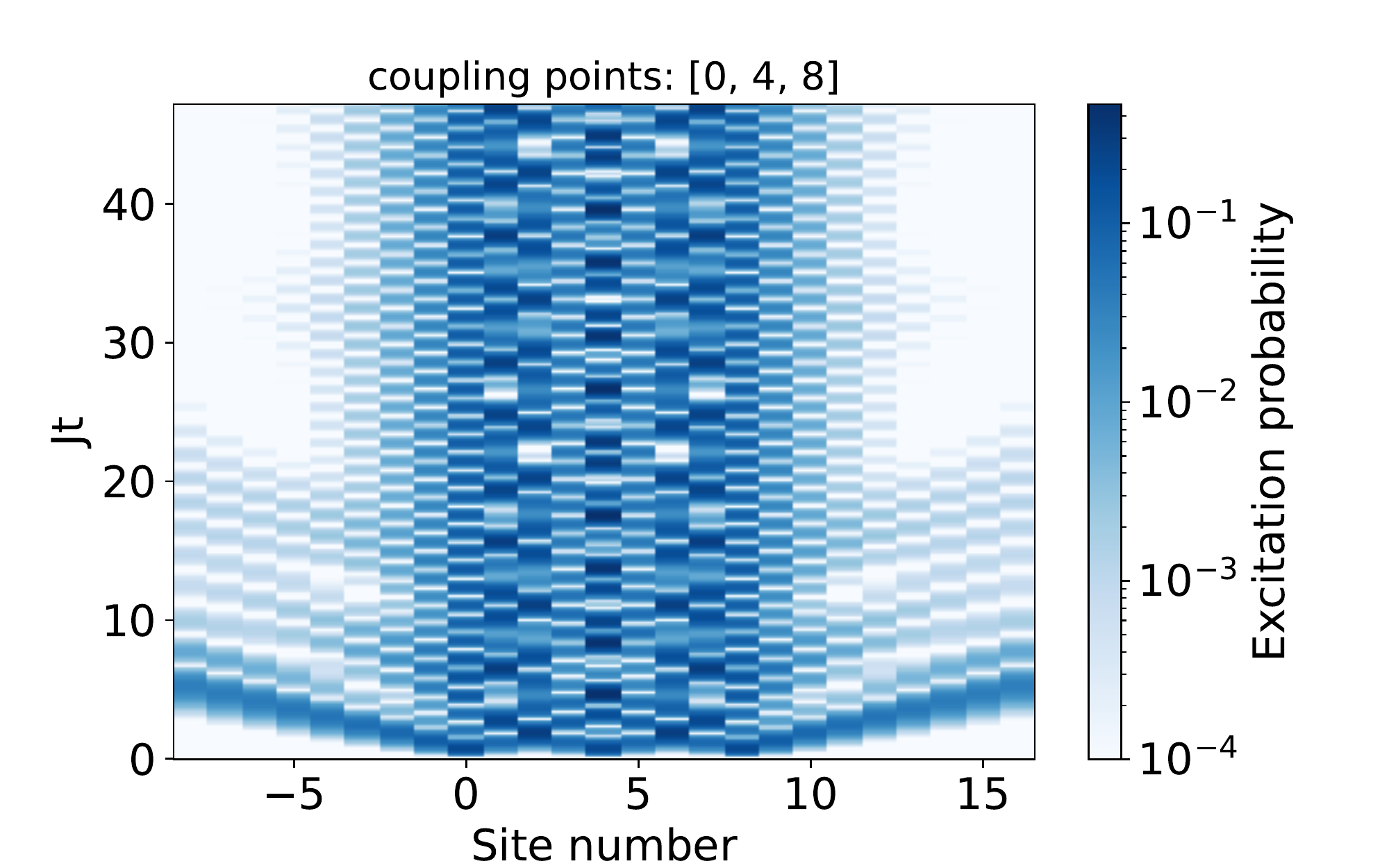}\hfill
  \subfigimg[width=0.48\textwidth, trim={0cm 0cm 1cm 0cm},clip]{(c)}{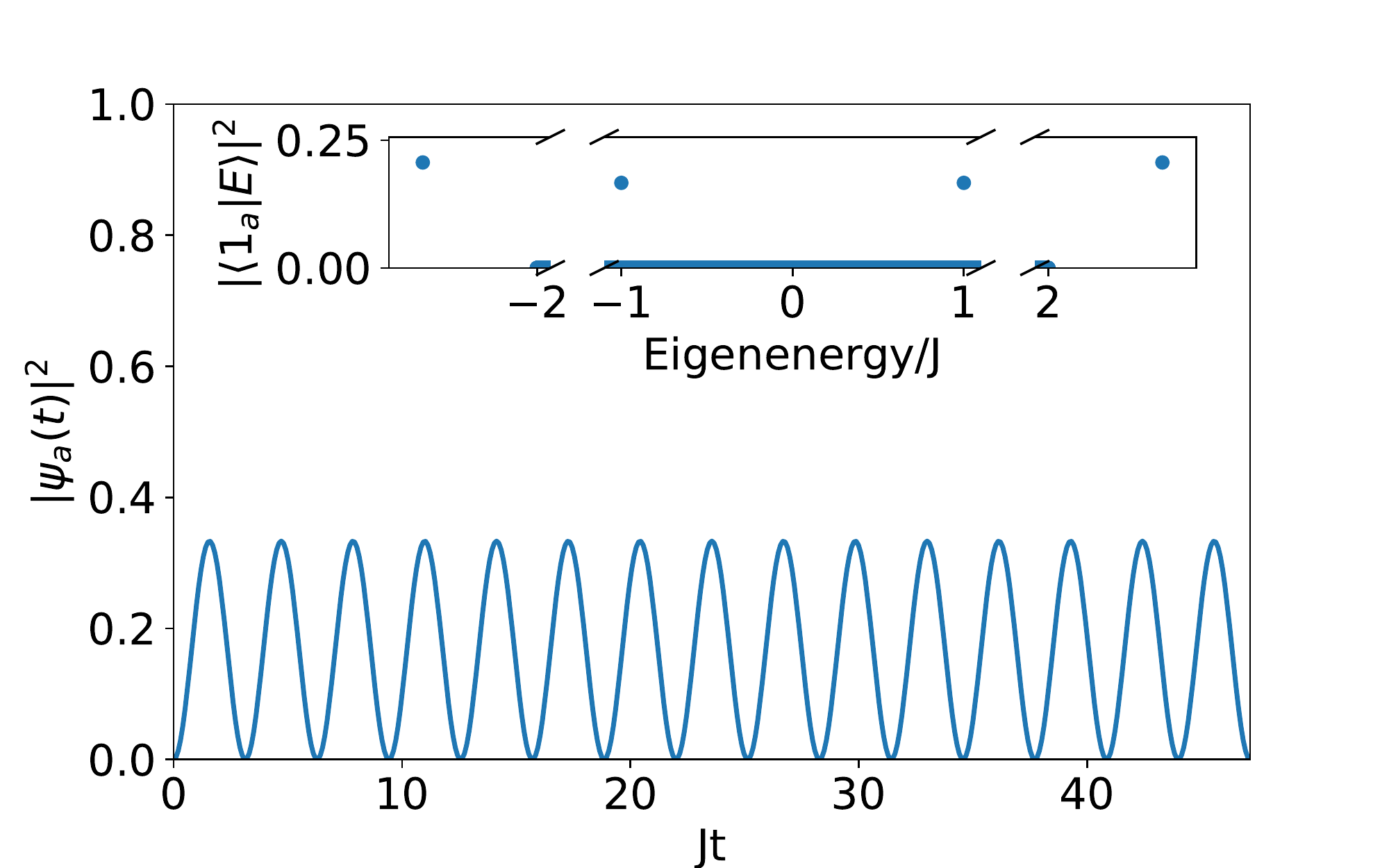}\hfill
  \subfigimg[width=0.48\textwidth, trim={0cm 0cm 1cm 0cm},clip]{(d)}{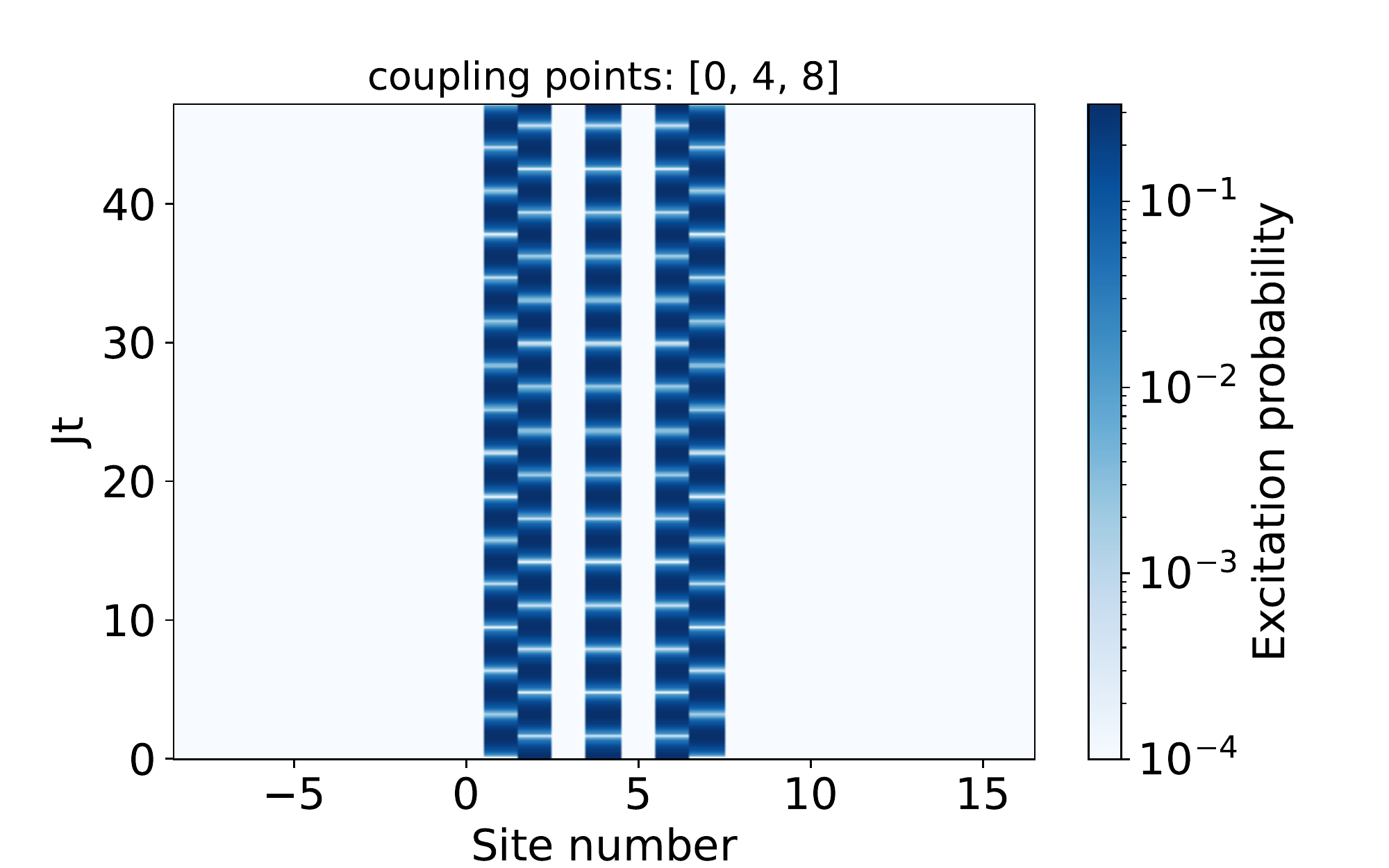}\hfill
  \caption[]{Similar to \figref{fig: M=3GAdecaySimulationResults}, in (a) and
  (b) we consider the case of giant atom decay into the 1D lattice chain but
  with $M = 3$ coupling points and $n_0 = 4$ sites between each coupling
  point.  In (c) and (d), we also consider the case where we have $M=3$
  coupling points and $n_0 = 4$ sites but with the giant atom lattice site
  initially in the vacuum state and the 1D lattice chain initialized in the
  state $\ket{p}$ according to \eqref{eqn: BIC subspace state p}.
  Remarkably, using \eqref{eqn: BIC rho0 M legs n0=4l}, the required giant
  atom coupling strength is $\rho_0/J=1$.  Due to the presence of the two
  BOCs with large emitter probabilities, for the case of the giant atom decay
  in (a) and (b), we see that the oscillation of the giant atom excitation
  probability in (a) is highly non-sinusoidal.  Moreover, in (b), we see that
  there is a leakage of photon excitation probability beyond the coupling
  points due to the presence of the BOC, which is non-ideal.  In (c) and (d),
  we see that because we are starting in the BIC subspace, the contributions
  due to the BOCs are totally eliminated.  We see perfectly sinusoidal
  oscillations of the giant atom excitation probability in (c), and we also
  see that the photon excitation probability is strictly confined to within
  the coupling points of the giant atom to the 1D lattice.  }
  \label{fig: M=3andn0=4BICsubspaceResults}
\end{figure*}

\begin{figure*}
  \centering
  \subfigimg[width=0.32\textwidth, trim={0cm 0cm 0cm 0cm},clip]{(a)}{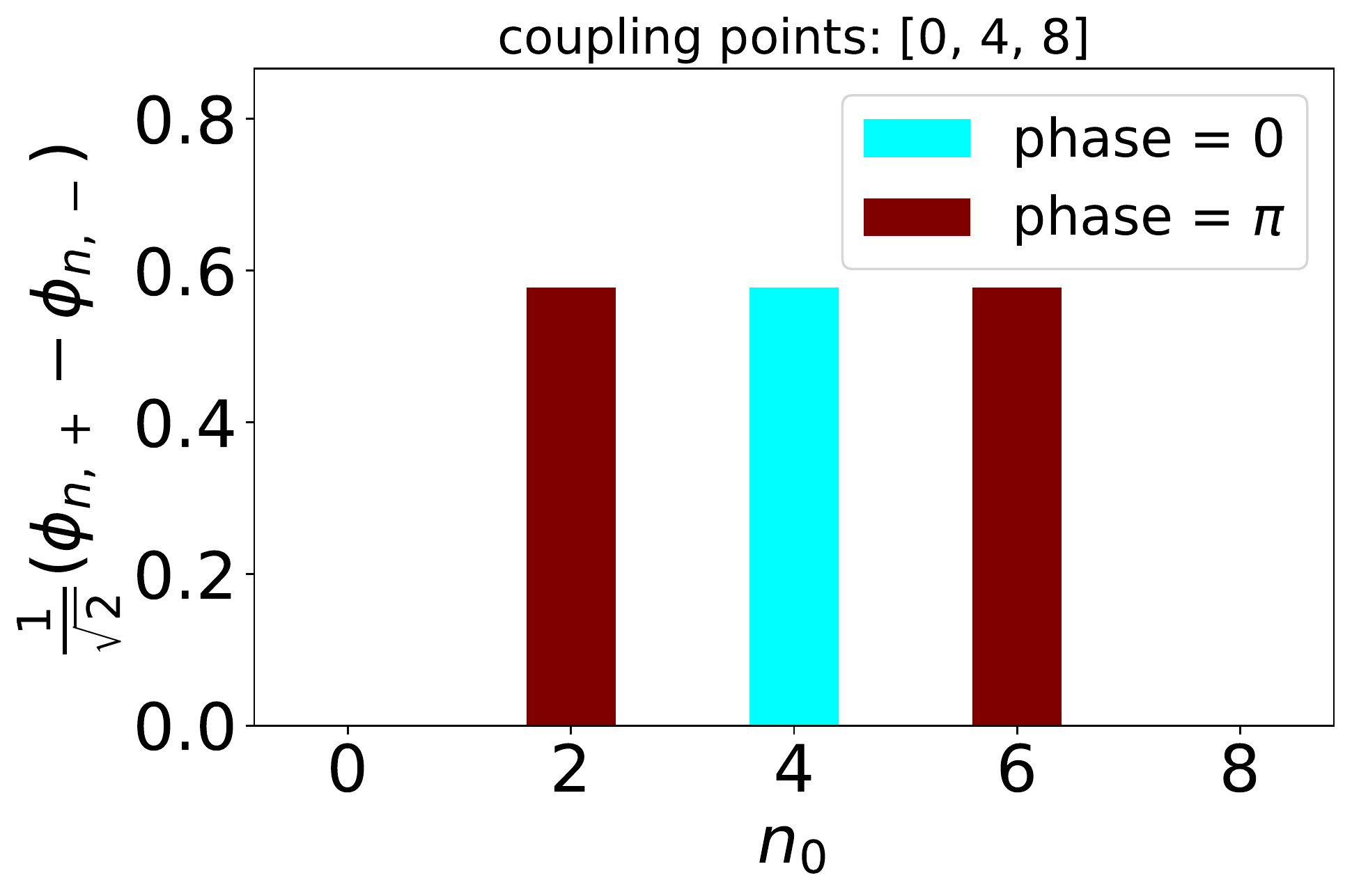}\hfill
  \subfigimg[width=0.32\textwidth, trim={0cm 0cm 0cm 0cm},clip]{(b)}{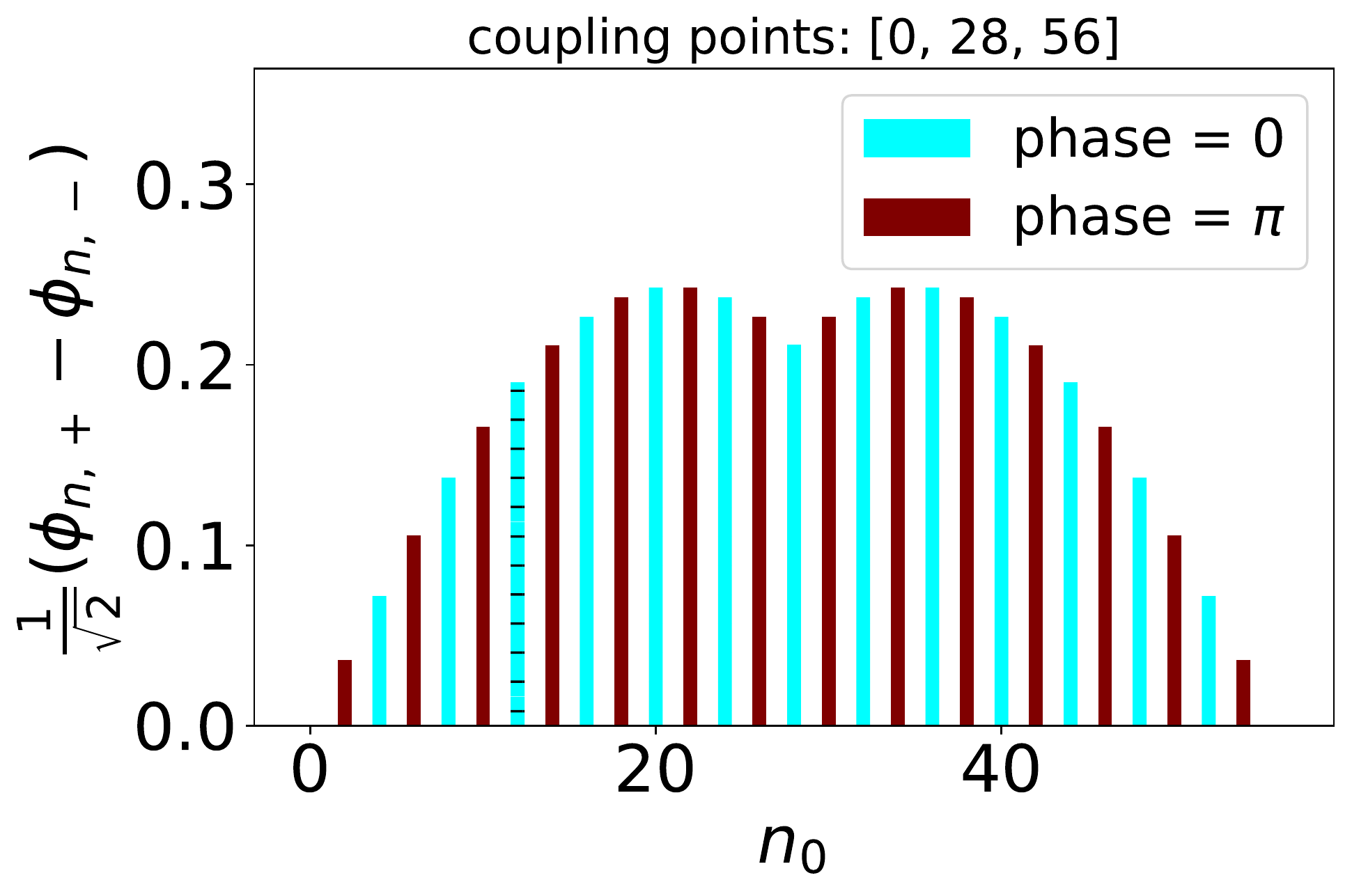}\hfill
  \subfigimg[width=0.32\textwidth, trim={0cm 0cm 0cm 0cm},clip]{(c)}{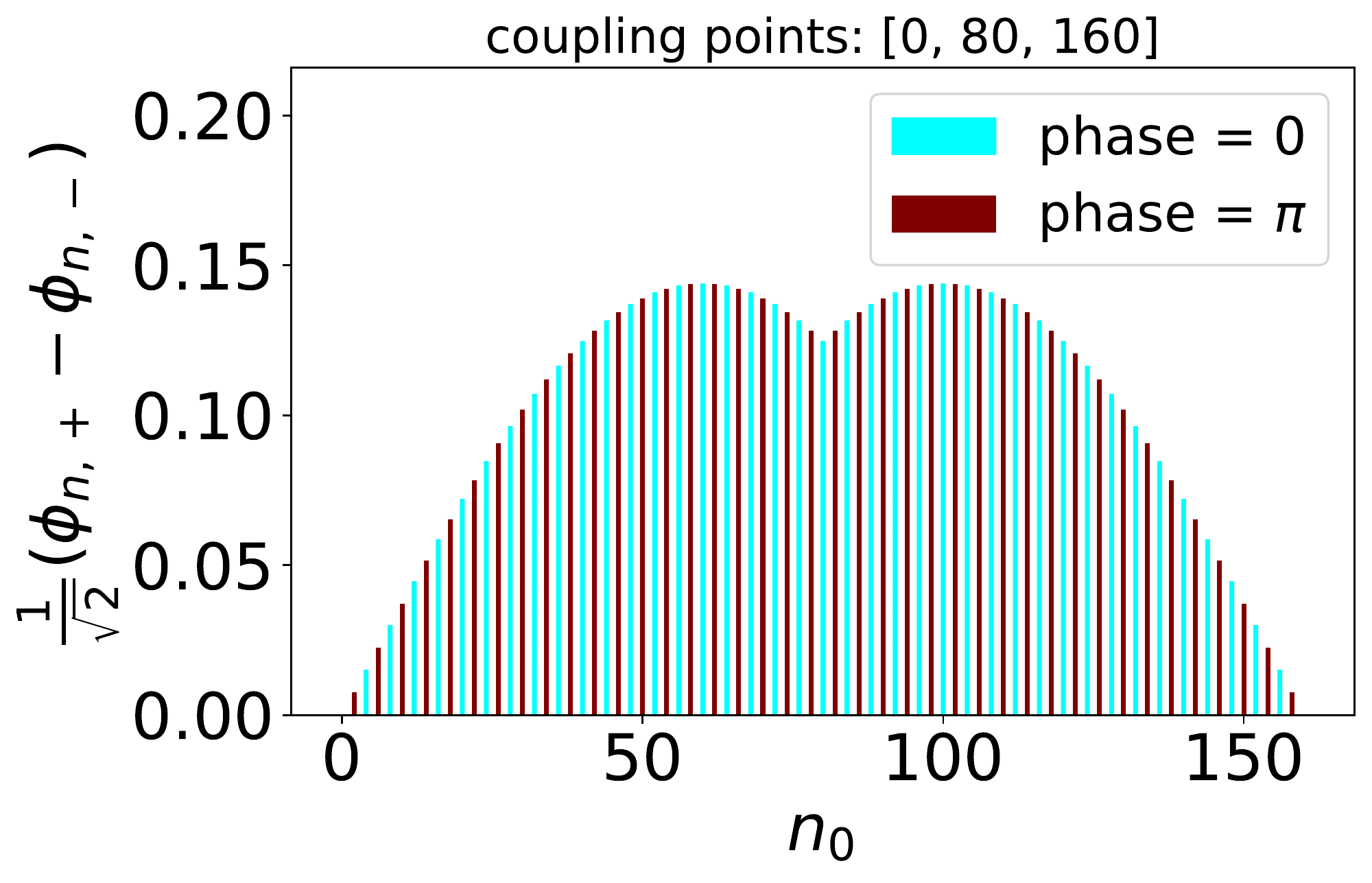}\hfill
  \caption[]{ Here we show the amplitudes and the phases of the photonic
  excitation required to initialize the 1D lattice chain in the state
  $\ket{p}$ given by \eqref{eqn: BIC subspace state p} for the case of $M=3$
  coupling points between the giant atom and the 1D lattice chain.  In
  (a),(b),(c), we plot the cases where we have $n_0=4$, $n_0=28$ and $n_0=80$
  respectively.  Sites with a phase of $0$ are indicated by a blue (lighter)
  color with a striped line within, whereas sites with a phase of $\pi$ are indicated by the a
  solid red (darker) color.}  
  \label{fig: BICsubspaceinitialization}
\end{figure*}

\begin{figure}[ht]
  \centering
  \subfigimg[width=0.48\textwidth, trim={0cm 0cm 1cm 0cm},clip]{(a)}{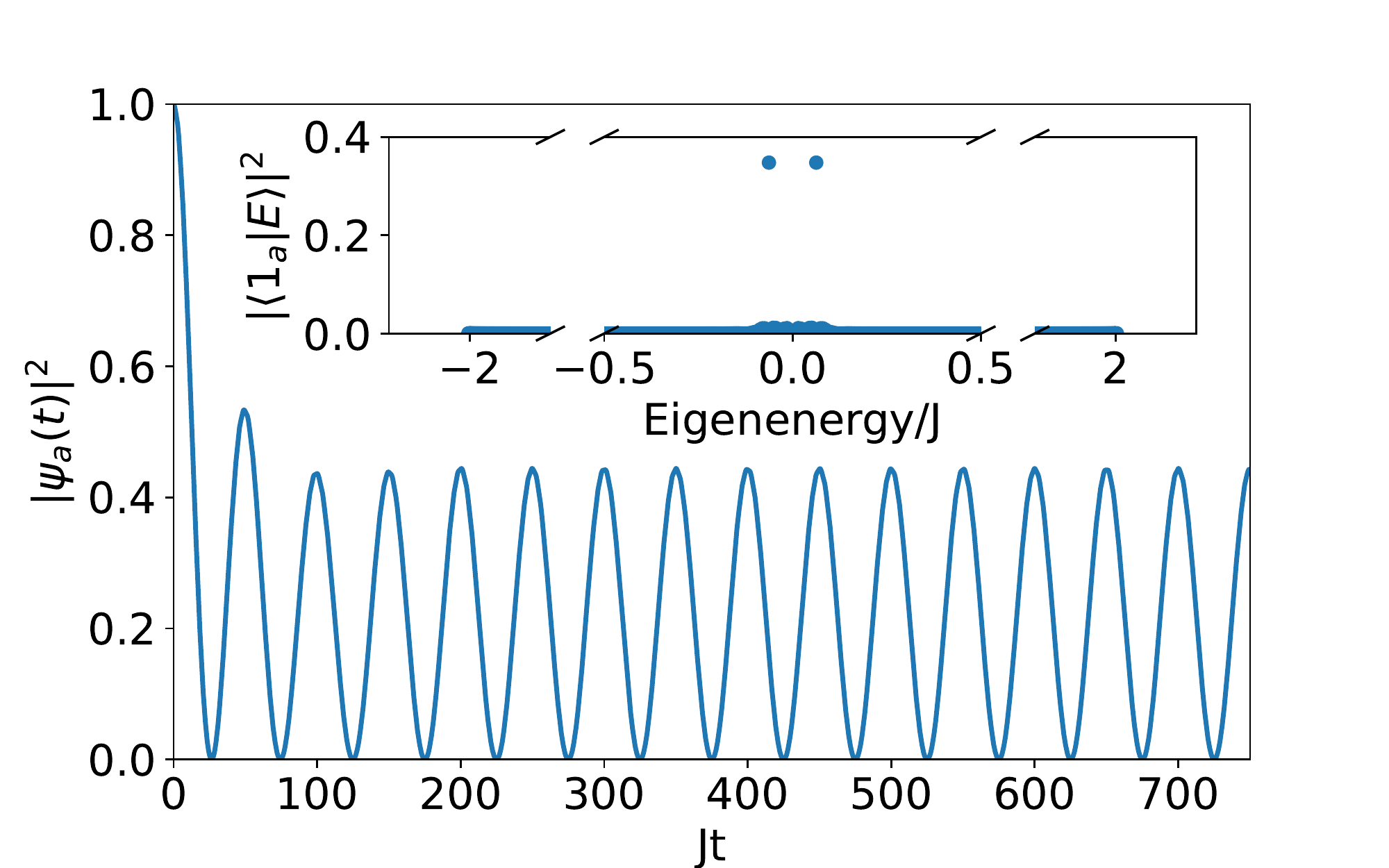}\hfill
  \subfigimg[width=0.48\textwidth, trim={0cm 0cm 1cm 0cm},clip]{(b)}{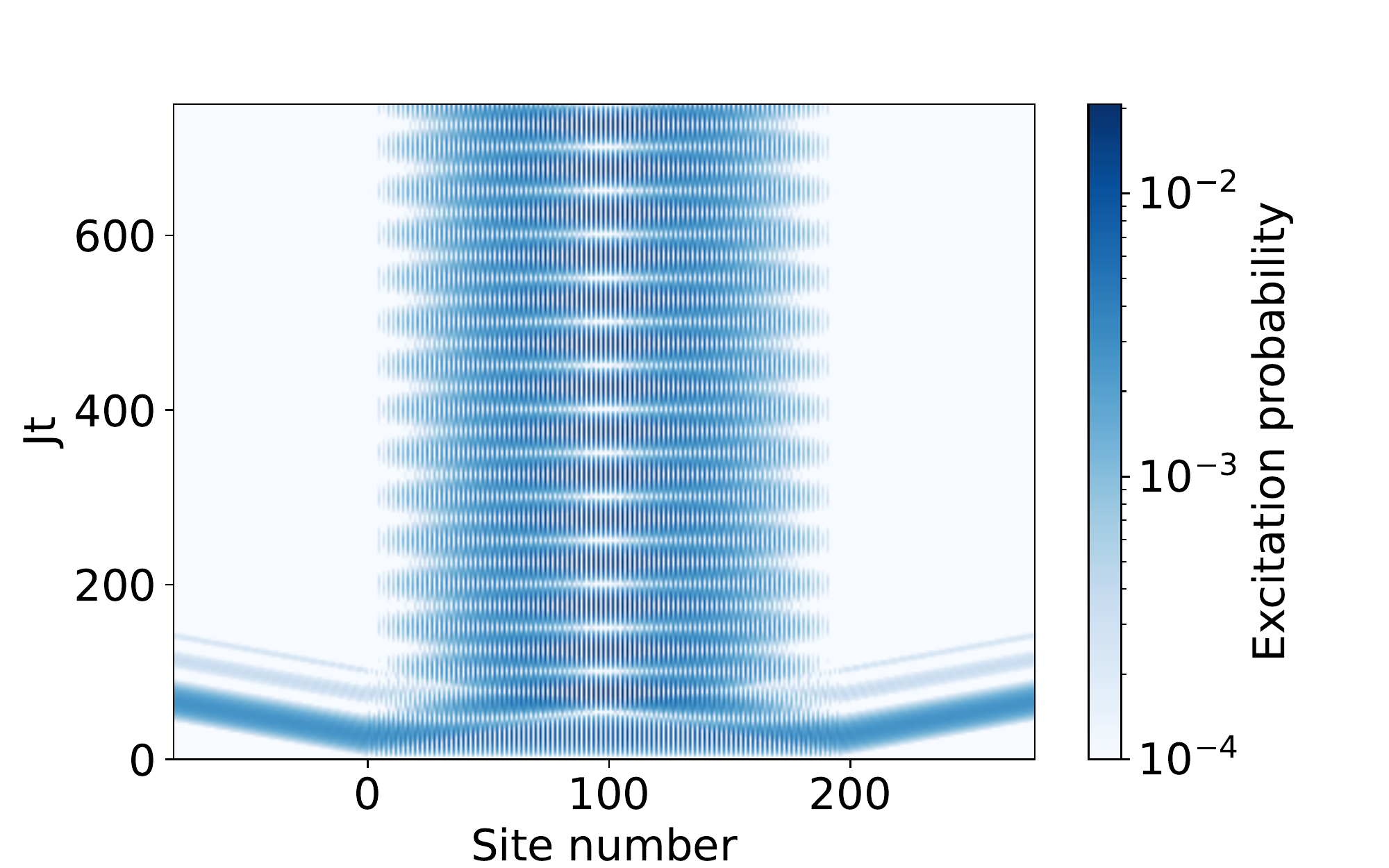}\hfill
  \caption[]{The analog of \figref{fig: M=3GAdecaySimulationResults} except
  that we have $M=50$ coupling points between the giant atom and the 1D
  photonic lattice for $n_0 = 4$.  As can be seen from the inset of (a), the
  emitter probabilities from the BOC are heaviliy suppressed as per
  \eqref{eqn: asymptotic BOC probability} which leads to perfectly sinusoidal
  oscillations of the excitation probability of the giant atom lattice site.
  We also see in (b) that the negligible BOC emitter probability leads to an
  absence of loss of photonic excitation probability into the 1D lattice
  chain.  By comparing (a) in this figure to \figref{fig:
  M=3GAdecaySimulationResults}, we see that a larger value of $M$ leads to
  larger amplitude oscillations in the excitation probability of the giant
  atom lattice site even for a comparatively small value of $n_0$, as per
  \eqref{eqn: BIC emitter probs M legs n0=4l}.  }
  \label{fig: M=50n0=4giantatomdecay}
\end{figure}

Here we first present in \figref{fig: M=3GAdecaySimulationResults} some
results for the $M=3$ giant atom decay coupled to a 1D photonic lattice with
various values of $n_0$.  For $M=3$, starting with a single excitation in the
giant atom, in the absence of imperfections due to the BOCs, we should expect
sinusoidal oscillations in the excitation probability of the giant atom
lattice site with oscillation amplitude $(2|\phi_a^{(\text{BIC})}|^2)^2
\approx 0.117411$, where we have used \eqref{eqn: asymptotic emitter prob as
n0=4l, n0 tends to infty} to obtain $|\phi_a^{(\text{BIC})}|^2 \approx 0.171$
for $M=3$.  However, for the $M=3$ case, as is seen in \figref{fig:
nonMarkovianityProbPlot}, the BOC emitter probabilities are actually quite
substantial, especially at small values of $n_0$.  Hence, this leads us to
consider a strategy for the $M=3$ case where instead of considering giant
atom decay, we initialize the lattice sites in the initial state \eqref{eqn:
BIC subspace state p} to eliminate the effects of the BOCs resulting in
complete storage of quantum information within the legs of the giant atoms,
and perfectly sinusoidal oscillations in the excitation probability of the
giant atom lattice site with oscillation amplitude
$2|\phi_a^{(\text{BIC})}|^2 \approx 0.33$.  An example for the $n_0=4$ case
is shown in \figref{fig: M=3andn0=4BICsubspaceResults}.  The amplitude and
phase of the initial photon excitation at each of the lattice sites can be
found using \eqref{eqn: BIC subspace state p}, where examples for various
values of $n_0$ are shown in \figref{fig: BICsubspaceinitialization}.
Finally, to show the effect of increasing $M$ on the quality of the giant
atom oscillating BIC, we plot the case of giant atom decay for $M=50$ and
$n_0=4$ case in \figref{fig: M=50n0=4giantatomdecay}.  We note that for this
value of $M$, we should expect the excitation probability in the giant atom
lattice site to oscillate with an amplitude of
$(2|\phi_a^{(\text{BIC})}|^2)^2 \approx 4/9$ as $|\phi_a^{(\text{BIC})}|^2$
approaches the asymptotic value of $1/3$ for increasing values of $M$.

\begin{figure}
  \centering
  \includegraphics[width=0.48\textwidth, trim={0.5cm 0cm 1.7cm 0cm},clip]{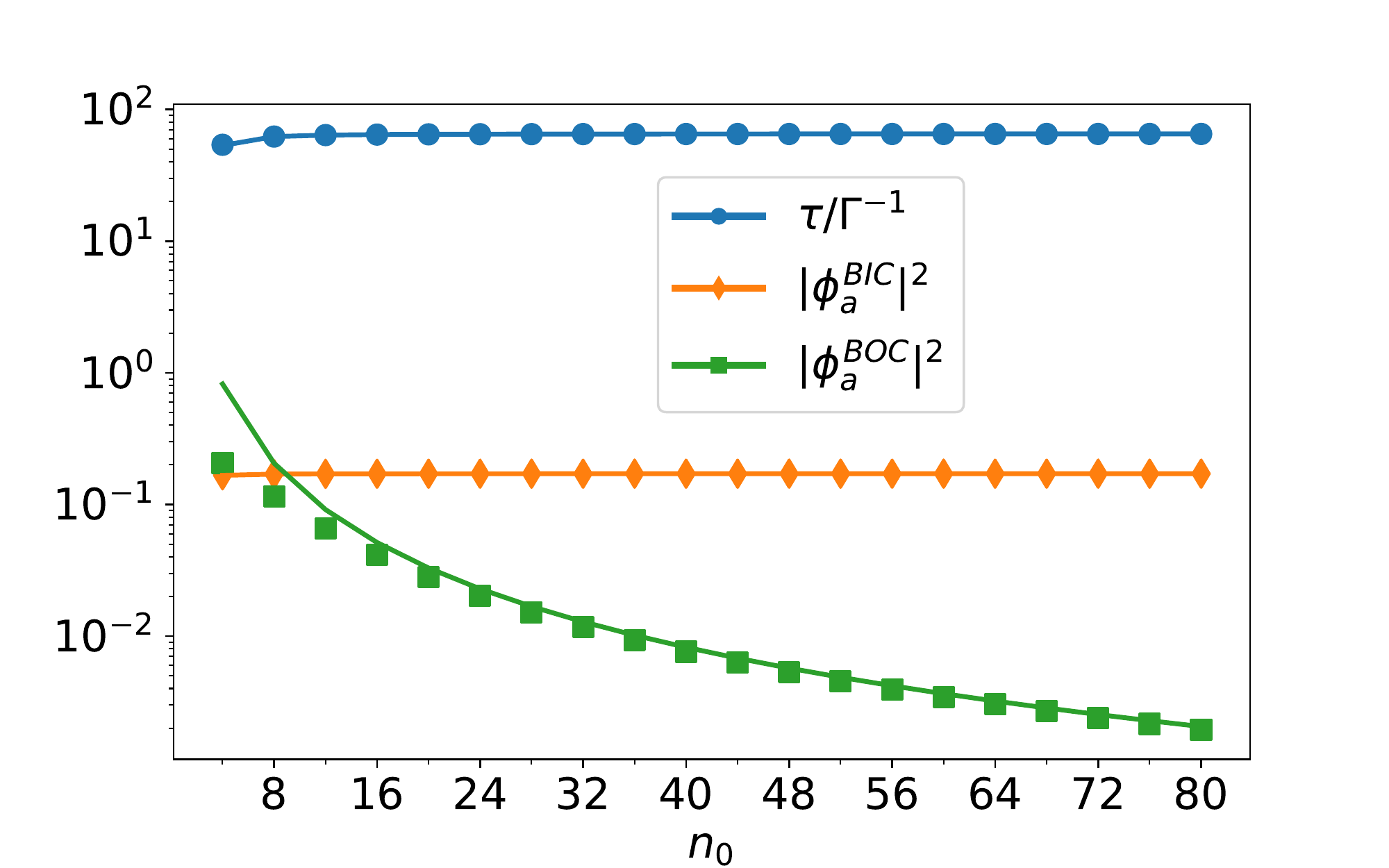}
  \caption[]{
  A plot showing how the emitter probabilities for the BICs, given by
  $|\phi_a^{(\text{BIC})}|^2$ and the BOCs, given by
  $|\phi_a^{(\text{BOC})}|^2$ scale as the number of sites $n_0$ between each
  coupling point increases.  On this figure we show the values computed
  numerically (scatter plot) as well as the values computed from our
  asymptotic expansions (continuous lines) as $n_0 \to \infty$ for the case
  where there are $M=3$ coupling points between the giant atom the the 1D
  lattice chain. On the same figure we also plot how $\tau/\Gamma^{-1}$
  scales with $n_0$.  As can be seen, as $\tau/\Gamma^{-1}$ increases and
  saturates at a value given by \eqref{eqn: nonMarkovianity asymptotic
  measure}, $|\phi_a^{(\text{BIC})}|^2$ also increases and saturates at a
  value given by \eqref{eqn: asymptotic emitter prob as n0=4l, n0 tends to
  infty}.  On the other hand, $|\phi_a^{(\text{BOC})}|^2$ decreases
  monotonically with increasing $n_0$.  We see that clearly, as the amount of
  non-Markovianity $\tau/\Gamma^{-1}$ in our system increases, we get better
  oscillating bound states.}
  \label{fig: nonMarkovianityProbPlot}
\end{figure}

\section{Experimental implementation}
\label{sec:experiments}
The Hamiltonian in \eqref{eqn: real space hamiltonian} can be simulated on a
variety of platforms, such as coupled cavity
arrays~\cite{hartmann_quantum_2008,majumdar_design_2012} and photonic
waveguide arrays~\cite{christodoulides_discretizing_2003,longhi_quantum_2009,
szameit_discrete_2010, aspuru-guzik_photonic_2010,garanovich_light_2012}.  In
the case of a photonic waveguide array, we would have one photonic waveguide,
which we call the giant atom waveguide, coupled to $M$ different photonic
waveguides that are already coupled to each other to form a linear chain of
$N$ waveguides, where the coupling is due to the evanescent field produced by
the photon propagating within the waveguide.  As the photon propagates in the
waveguide, we have the relation $z = ct$, where $c$ is the group velocity of the
photon in the waveguide and $z$ is the distance along the waveguide that the
photon has propagated for.

\begin{figure}[h]
  \centering
  \includegraphics[width=0.48\textwidth, trim={0cm 0cm 9cm 0cm},clip]{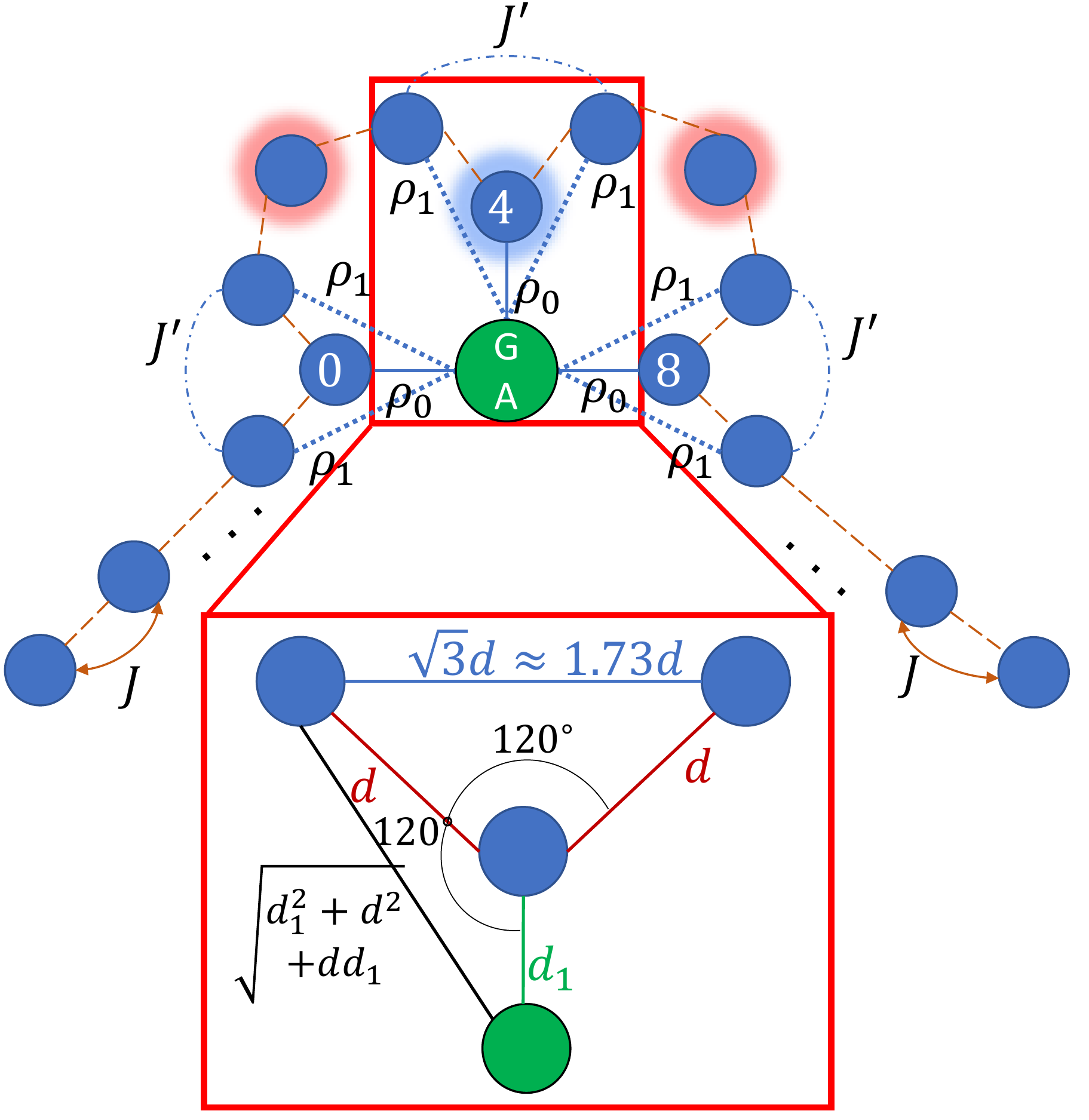}
  \caption[]{
  Possible experimental setup to see oscillating BICs.  Here, we have $M=3$
  and $n_0=4$.  The giant atom waveguide is in green, with the letters ``GA''
  inside whereas the waveguides in the waveguide chain are in blue.  The
  coupling strength $\rho_0$ can be found from \eqref{eqn: BIC rho0 M legs
  n0=4l}.  Here, $\rho_1$ is the coupling between the giant atom and the
  next-nearest neighbor sites and $J^\prime$ is the hopping strength between
  the sites that are $\pm1$ from the coupling point.  In this set-up, we
  start with the giant atom in the vacuum state and instead initialize the
  rest of the waveguide chain in the state $\ket{p}$ according to \eqref{eqn:
  BIC subspace state p}.  This means that as per \figref{fig:
  BICsubspaceinitialization} for the case where $n_0 = 4$, we need to send in
  a photon with equal probability amplitude in waveguides $2$, $4$, $6$ which
  we have highlighted.  The red glow (site 2,6) indicates a phase of $\pi$
  for the photon, whereas a waveguide with blue glow (site 4) indicate a
  phase of $0$.  For the array of photonic waveguides, the coupling strength
  between each waveguide decays exponentially with the distance between the
  waveguides \cite{jiao_two-dimensional_2021}, hence by choosing the
  distances $d, d_1$ in the experimental setup appropriately as per the
  inset, we can obtain $\rho_1 \ll \rho_0$ and $J^\prime \ll J$ which allow
  us to neglect the next-nearest neighbor interactions.}
  \label{fig: proposed setup for oscillating BS BIC subspace}
\end{figure}

Following our formalism above, the nearest-neighbour coupling of the photonic
waveguides in the linear chain with coupling strength $J$ gives us the tight
binding Hamiltonian $H_{\text{wg}}$, whereas the coupling between the giant
atom waveguide and the linear chain of waveguides at $M$ different points,
each spaced $n_0$ apart, with coupling strength $\rho_0$ gives us the
interaction Hamiltonian $H_{\text{int}}$.  Taking the constraints of current
experimental capabilities in mind, we propose an experimental setup for the
case where $M=3$ and $n_0 = 4$ using the BIC subspace initialization in
\figref{fig: proposed setup for oscillating BS BIC subspace}.  For this
photonic waveguide array system, the BIC initialization according to
\figref{fig: proposed setup for oscillating BS BIC subspace} can be achieved
deterministically with a spatial light modulator that modulates a single
photon source \cite{tentrup_transmitting_2017}.  Alternatively, one can also
prepare the oscillating BIC probabilistically by initializing an excitation
only in the giant atom waveguide and perform photodetection on the sites
outside of the giant atom coupling points, and postselect on the no-detection
events.

In \figref{fig: proposed setup for oscillating BS BIC subspace} we have also
denoted the next-nearest neighbor coupling between the giant atom waveguide
and the sites $0 \pm 1$, $n_0 \pm 1$ and $2n_0 \pm 1$ with $\rho_1$, and also
the next-nearest neighbor hopping between the lattice sites $0 \pm 1$, $n_0
\pm 1$ and $2n_0 \pm 1$ with $J^\prime$.  In general, the presence of
$J^\prime$ and $\rho_1$ are unwanted imperfections, yet we note that by
choosing the geometry and the distances accordingly as per the inset in
\figref{fig: proposed setup for oscillating BS BIC subspace}, we can minimize
the contributions from $\rho_1$ as well as $J^\prime$.  To do so, we first
use \eqref{eqn: BIC energies M legs n0=4l} to calculate the emitter energies
for $n_0=4$, which would give us the oscillation period $T = \pi$ for the
oscillating BIC.  This means that to see an appreciable number of
oscillations, we could simulate up to $Jz = 5T \approx 15$.  Now, suppose
that experimentally, we can only have photonic waveguides with length
$z_{\text{max}}$.  This means that we require $J = 5T/z_{\text{max}}$.
Henceforth, we shall assume $z_{\text{max}} = 100$ mm which has been done
experimentally before~\cite{tang_generating_2022}.  From \eqref{eqn: BIC rho0
M legs n0=4l}, we obtain $\rho_0/J=1$ which tells us to set $\rho_0=J$.  It
is known that the evanescent coupling strength between waveguides decay
exponentially with the distance between
them~\cite{jiao_two-dimensional_2021}.  Using the experimental values
obtained in~\cite{jiao_two-dimensional_2021} for the aforementioned
exponential relationship between coupling strength and distance, together
with the geometry of the proposed setup in the inset of \figref{fig: proposed
setup for oscillating BS BIC subspace}, we obtain $\rho_0=J=0.15 \,
\text{mm}^{-1}$, $\rho_1 = 0.0286 \rho_0$ and $J^\prime = 0.0286J$ which is
nearly negligible.  Thus, our proposed oscillating BICs are experimentally
feasible using state-of-the-art photonic waveguide arrays.

\begin{figure*}[t]
  \centering
  \subfigimg[width=0.48\textwidth, trim={0cm 0cm 1cm 0cm},clip]{(a)}{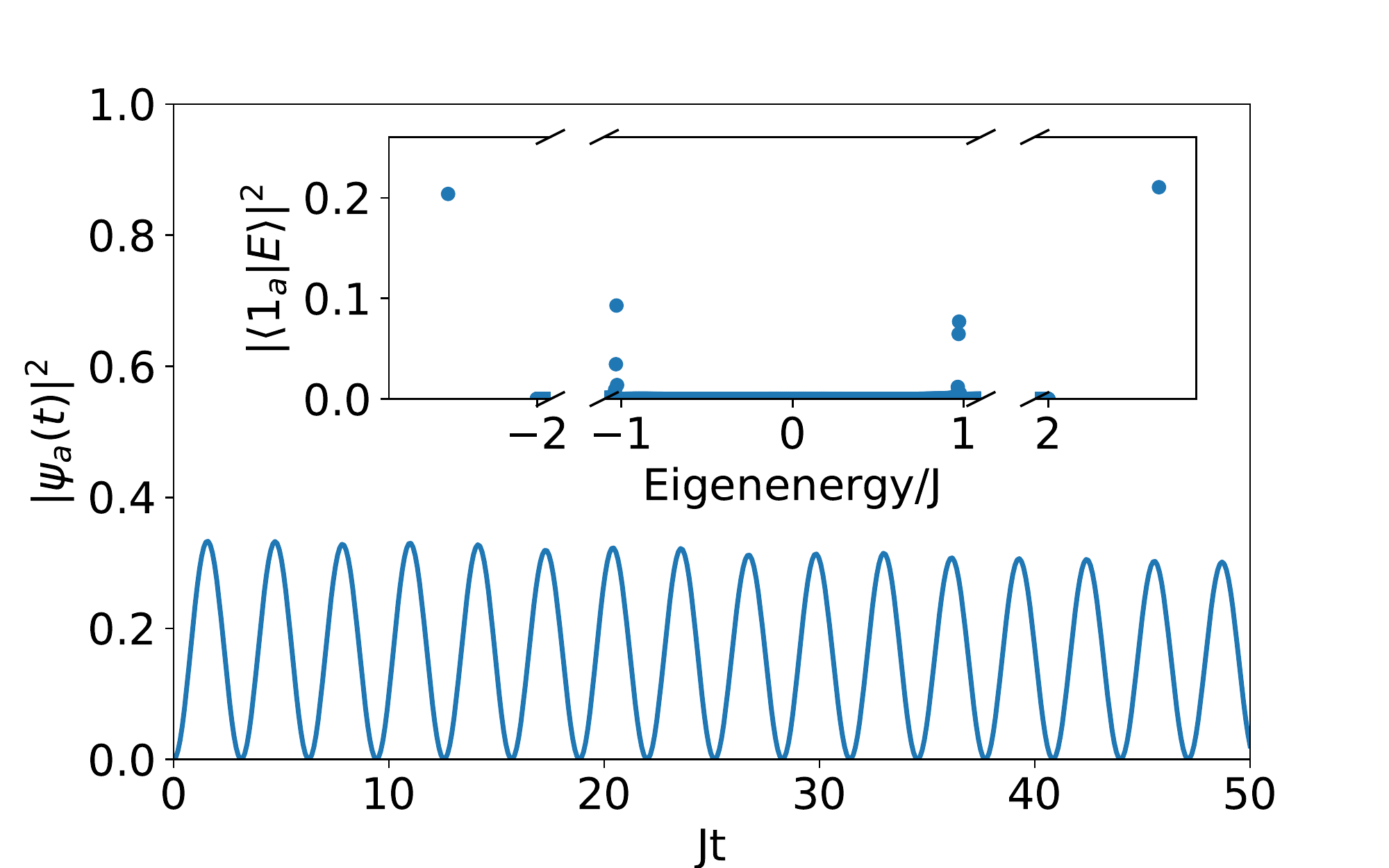}\hfill
  \subfigimg[width=0.48\textwidth, trim={0cm 0cm 1cm 0cm},clip]{(b)}{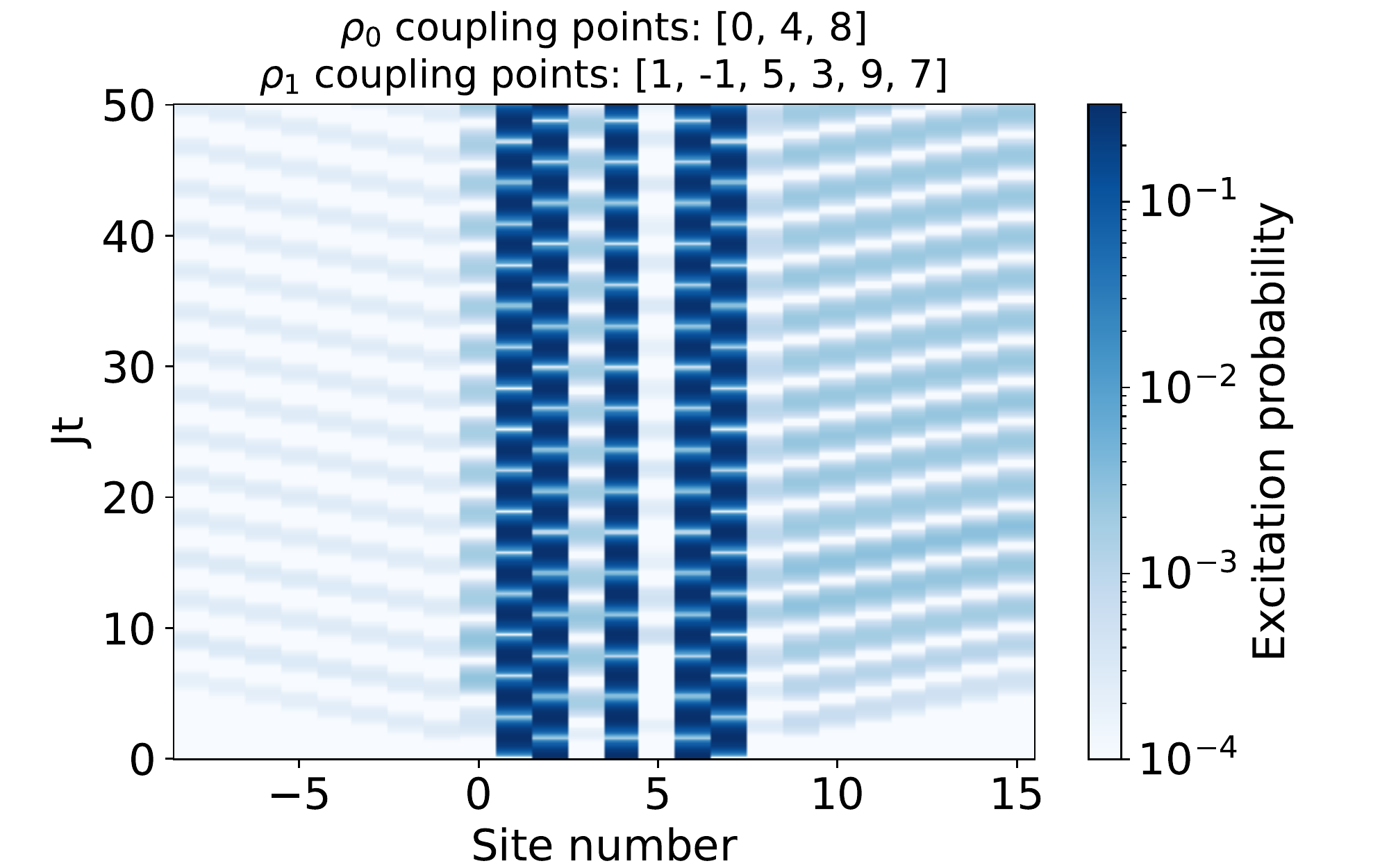}\hfill
  \caption[]{ 
  Simulation results for the proposed experimental implementation with the
  setup in \figref{fig: proposed setup for oscillating BS BIC subspace}.
  Here, we initialize the waveguide array in the state $\ket{p}$ and with the
  giant atom lattice site in the vacuum state.  Also, with reference to
  \figref{fig: proposed setup for oscillating BS BIC subspace}, we consider
  the case where we have $\rho_1 = 0.0286 \rho_0$ and $J^\prime = 0.0286J$ to
  show the impact of these experimental imperfections on our simulation.  The
  corresponding ideal case for which $\rho_1 = J^\prime = 0$ can be found in (c)
  and (d) of \figref{fig: M=3andn0=4BICsubspaceResults}.  By comparing the
  ideal case with (a) and (b) in this figure, we see that the presence of
  experimental imperfections leads to a leakage of photon excitation
  probability beyond the coupling points of the giant atom, yet the effect is
  nearly negligible up to $Jz = 15$, which is a large enough value of $Jz$ to
  see an appreciable number of oscillations.}
  \label{fig: experimental implementation simulation}
\end{figure*}

Simulation results for $M=3$ and $n_0=4$ when $\rho_1=0$ and $J^\prime=0$ can
be found in \figref{fig: M=3andn0=4BICsubspaceResults}.  The corresponding
results when $\rho_1 = 0.0286 \rho_0$ and $J^\prime = 0.0286J$ can be found in \figref{fig: experimental implementation simulation}.

\section{Conclusion}
\label{sec:conclusion}
In this paper, we study the phenomenon of oscillating BIC in a discrete 1D photonic lattice using a single emitter coupled to multiple
lattice sites, which can be considered as the
discrete analog of a giant atom coupled to a continuous waveguide. The key difference between our work and the oscillating BICs found in continuous waveguide systems~\cite{guo_oscillating_2020} is the presence of a finite energy band, which contributes band-edge effects to the giant atom dynamics. This gives us new conditions for the existence of oscillating BICs which lead to persistent oscillations of energy between the coupling points of the giant atom to the 1D lattice. The presence of bound states outside the continuum (BOC) hinders the trapping of excitation between the giant atom coupling points and is detrimental to the sinusoidal oscillations in the giant atom probability. Crucially, we find that these unwanted BOCs can be suppressed drastically by increasing either the number of coupling points $M$ or the number of lattice sites $n_0$ between each coupling point, with the BOC contribution scaling as $1/M^2$ and $1/n_0^2$. With this, we can summarize our key results for the conditions to produce optimal oscillating BICs to be: (1) $n_0 = 4l, l \in \mathbb{Z}^+$ sites between each coupling point, (2) Large $n_0$ and (3) Large M. In practice however, we find that a moderate $M$ and $n_0$ suffice to achieve good oscillating BICs with significant giant atom probability. Alternatively, by initializing the lattice sites in the BIC subspace which we have calculated, the BOC contributes can be completely eliminated, resulting in perfect oscillating BICs even for small $M$ and $n_0$. We
stress that the oscillating BIC in our system is inherently a non-Markovian phenomenon due to the significant propagation time between the giant atom coupling points compared to the relaxation timescale of the giant atom. Moreover, we show that as the
non-Markovianity in our system increases, the oscillation amplitude of the BICs increases, improving the storage of quantum information within the coupling points. To illustrate the feasibility of our theoretical model, we propose an experimental implementation of our
system on photonic waveguide arrays and show that our oscillating BICs can be practically achieved even with current experimental limitations.

Our work provides a firm theoretical basis for oscillating BICs in discrete systems. In particular, oscillating BICs in discrete systems
offer new possibilities that cannot be replicated in continuous systems,
such as the ability to initialize the system in the BIC subspace by simply controlling the amplitude and phase of
the excitation at particular lattice sites. This allows us to achieve long-time storage of quantum information within the confines of the giant atom coupling points, limited only by the intrinsic coherence time of the photonic lattice. Our setup can also be regarded as an effective cavity, serving as a physical implementation of non-Markovian cavity-QED setups~\cite{crowder2020quantum,regidor2021cavitylike,guimond2016rabi,cernotik2019cavity,du2021single}. While we have
considered the tight-binding dispersion relation in this work, the phenomenon of oscillating BICs can be generally observed in discrete systems with other dispersion relations, which can be considered for future work. Another promising direction is to study the oscillating BIC phenomenon in higher-dimensional lattices~\cite{gonzalez2019engineering} or in synthetic dimensions~\cite{du2022giant,xiao2022bound}.  

\section*{Acknowledgements}
K.H.L, W.K.M and L.C.K. are grateful to the National Research Foundation, Singapore and the Ministry of
Education, Singapore for financial support. The authors thank Anton Frisk Kockum for useful discussions.


%

\onecolumngrid
\newpage
\appendix

\section{Writing \eqref{eqn: real space hamiltonian} in
\texorpdfstring{$k$}{k}-space}
\label{Appendix: k-space hamiltonian derivation}
From \eqref{eqn: k-space annihilation operators}, we can express
$H_{\text{wg}}$ as
\begin{align}
    H_{\text{wg}}= & J\sum_{n}b_{n}^{\dagger}b_{n+1}+\text{H.c}\nonumber \\ 
    = & \frac{J}{2\pi}\iint
    dk\,dk^{\prime}\sum_{n}e^{i(k^{\prime}-k)n}e^{ik^{\prime}}c^{\dagger}(k)c(k^{\prime})+\frac{J}{2\pi}\iint
    dk\,dk^{\prime}\sum_{n}e^{-i(k^{\prime}-k)n}e^{-ik^{\prime}}c^{\dagger}(k^{\prime})c(k)\nonumber
    \\
    = & \frac{J}{2\pi}\iint
    dk\,dk^{\prime}\sum_{n}e^{i(k^{\prime}-k)n}e^{ik^{\prime}}c^{\dagger}(k)c(k^{\prime})+\frac{J}{2\pi}\iint
    dk^{\prime}\,dk\sum_{n}e^{-i(k-k^{\prime})n}e^{-ik}c^{\dagger}(k)c(k^{\prime})\nonumber
    \\
    = & J\iint
    dk\,dk^{\prime}\left(\frac{1}{2\pi}\sum_{n}e^{i(k^{\prime}-k)n}\right)\left(e^{ik^{\prime}}+e^{-ik}\right)c^{\dagger}(k)c(k^{\prime})\nonumber
    \\
    = & J\iint
    dk\,dk^{\prime}\delta(k^{\prime}-k)\left(e^{ik^{\prime}}+e^{-ik}\right)c^{\dagger}(k)c(k^{\prime})\nonumber
    \\
    = & J\int dk\,\left(e^{ik}+e^{-ik}\right)c^{\dagger}(k)c(k)\nonumber \\
    = & \int dk\,2J\cos(k)c^{\dagger}(k)c(k)\nonumber \\
    = & \int dk\,\omega(k)c^{\dagger}(k)c(k),
\end{align}
where we have defined $\omega(k)\equiv2J\cos(k)$.  Here, we have used the
fact that for $(k^\prime - k) \in [0,2\pi)$,
$\frac{1}{2\pi}\sum_{n}e^{i(k^{\prime}-k)n} = \delta(k^\prime - k)$.
Similarly, we can also express $H_{\text{int}}$ as
\begin{align}
    H_{\text{int}} &= \sum_{j=1}^{k} \rho_{j}(a^\dag b_{n_j} + \text{H.c}) \nonumber \\
    &= \sum_{j=1}^{k} \rho_{j}\left\{a^\dag
    \left(\frac{1}{\sqrt{2\pi}}\int_{-\pi}^\pi dk\, e^{ikn_j}c(k)\right) +
    \text{H.c}\right\} \nonumber \\
    &= \int_{-\pi}^\pi dk\,\left\{\left(\frac{1}{\sqrt{2\pi}}\sum_{j=1}^k \rho_j
    e^{ikn_j}\right)a^\dagger c(k) + \text{H.c} \right\} \nonumber \\
    &\equiv \int_{-\pi}^\pi dk\,\left\{G(k)a^\dagger c(k) + \text{H.c} \right\},
\end{align}
where we have defined $G(k) = \left(\frac{1}{\sqrt{2\pi}}\sum_{j=1}^{K} \rho_j
e^{ikn_j}\right)$.

\section{Derivation of conditions for BIC}
\label{Appendix: Omega equation derivation}
First, to derive \eqref{eqn: Omega must be in band} and the condition for the
BIC to have a finite norm , we shall first write our energy eigenstate in the
one-excitation subspace as
\begin{equation}
    \ket{E} = \phi_a \ket{1_a} + \int dk \,e(k) \ket{1_k}
\end{equation}
where $\phi_a = \braket{1_a}{E}$ and $e(k) = \braket{1_k}{E}$.  Thereafter, we
consider the energy eigenvalue equation $H\ket{E} = \Omega \ket{E}$.  To
evaluate $H\ket{E}$, we use the commutation relations $[a,a^\dagger] = 1$ and
$[c(k),c^\dagger(k^\prime)] = \delta(k-k^\prime)$, and finally we arrive at
\begin{equation}
    H\ket{E} = \left(\omega_a \phi_a + \int dk\,G(k)e(k)\right)\ket{1_a} + \int
    dk\,\left(\omega(k)e(k) + G^*(k)\phi_a\right)\ket{1_k}.
\end{equation}
Comparing the above equation with $\Omega\ket{E} = \Omega (\phi_a \ket{1_a} + \int dk \,e(k) \ket{1_k})$, we arrive at two simultaneous equations 
\begin{subequations}
    \label{eqn: one excitation subspace energy eigval eqn simultaneous eqn}
    \begin{align}
        \Omega \phi_a &= \omega_a \phi_a + \int dk\, G(k)e(k) \\ 
        \label{eqn: finite norm condition}
        \Omega e(k) &= \omega(k)e(k) + G^*(k)\phi_a
    \end{align}
\end{subequations}
which we can then solve to obtain \eqref{eqn: Omega must be in band}.  Next,
we note that in the one-excitation subspace, we have
\begin{align}
  \braket{E}{E} &= |\phi_a|^2 + \int dk \, |e(k)|^2 \nonumber \\
  &= |\phi_a|^2 \left(1 + \int dk\, \frac{|G(k)|^2}{(\Omega-\omega(k))^2}\right)
  \nonumber \\
  &= |\phi_a|^2 \left(1 + \int d\omega\, \frac{\rho(\omega)|G(k(\omega))|^2}{(\Omega-\omega)^2}\right)
\end{align}
where in the second line, we used \eqref{eqn: finite norm condition} and in
the third line, we used $\rho(\omega) \equiv \frac{\partial k}{\partial
\omega}$.  Hence we see that the requirement that $\ket{E}$ has a finite norm
corresponds to $\rho(\omega)|G(k(\omega))|^2$ vanishing at least as fast as
$\sim(\Omega-\omega)^2$ as $\omega \to \Omega$. 

Next, to derive \eqref{eqn: detuning in terms of Sigma(s)}, we note that 
\begin{align}
    \Sigma(s=-i\Omega \pm 0^+) &= \int_{-\pi}^\pi dk \frac{|G(k)|^2}{\Omega -
    \omega(k)} \nonumber \\
    &= \int_{-2J}^{2J} d\omega \frac{\rho(\omega)|G(k(\omega))|^2}{\Omega -
    \omega} \nonumber \\
    &= \lim_{\epsilon \to
    0}\left(\int_{-2J}^{\Omega-\epsilon}d\omega\,\frac{\rho(\omega)|G(k(\omega))|^2}{\Omega
    - \omega} + \int_{\Omega +
    \epsilon}^{2J}d\omega\,\frac{\rho(\omega)|G(k(\omega))|^2}{\Omega - \omega}
    \right) + \int_C  d\omega\,\frac{\rho(\omega)|G(k(\omega))|^2}{\Omega - \omega} \nonumber \\
    &= \mathcal{P}\left(\int_{-2J}^{2J} d\omega \frac{\rho(\omega)|G(k(\omega))|^2}{\Omega -
    \omega}\right) \mp i \pi \rho(\Omega)|G(k(\Omega))|^2
\end{align}
where in the second last line, we note that the contour $C$ is a
semi-circular contour either in the top half or bottom half of the complex
plane, depending on the sign of $\pm 0^+$, and in the last line we used the
residue theorem for the special case of a semi-circular contour.
$\mathcal{P}()$ here denotes the Cauchy Principal value of the integral
enclosed in the parentheses.  Taking the real part of the last line,
\eqref{eqn: detuning in terms of Sigma(s)} follows immediately.  We note here
that if $\Omega$ also fulfils \eqref{eqn: coupling to continuum vanishes},
then the second term vanishes, which gives us
\begin{equation}
    \Sigma(s=-i\Omega \pm 0^+) = \text{Re}[\Sigma(s=-i\Omega \pm 0^+)].
\end{equation}

\section{Decay dynamics}
\label{Appendix: decay dynamics derivation}
Here, we derive \eqref{eqn: decay dynamics}.  By writing down the
Schrodinger equation $H\ket{\psi(t)} = i \frac{\partial}{\partial
t}\ket{\psi(t)}$ in the one-excitation subspace as per \eqref{eqn: one
excitation subspace ansatz}, we arrive at two coupled equations 
\begin{subequations}
    \begin{align}
        \label{eqn: giant atom eqn from schrodinger eqn}
        i\frac{\partial \psi_a(t)}{\partial t} &= \omega_a \psi_a(t) + \int_{-\pi}^\pi G(k)\psi(k,t)dk \\
        \label{eqn: waveguide chain eqn from schrodinger eqn}
        i\frac{\partial \psi(k,t)}{\partial t} &= \omega(k)\psi(k,t) + G^*(k)\psi_a(t).
    \end{align}
\end{subequations}
To solve the two above equations for $\psi_a(t)$, we follow the standard
procedure of first integrating the ``bath'' equation, which is \eqref{eqn:
waveguide chain eqn from schrodinger eqn} in our case, to get
\begin{equation}
    \psi(k,t) = -i G^*(k) e^{-i\omega(k)t}\int_0^t dt^\prime e^{i\omega(k)t^\prime}\psi_a(t^\prime) + e^{-i\omega(k)t}\psi(k,0).
\end{equation}
Then, we substitute the above equation into \eqref{eqn: giant atom eqn from
schrodinger eqn} to get
\begin{equation}
  \frac{\partial \psi_a(t)}{\partial t} = -i\omega_a \psi_a(t) - \int_{-\pi}^\pi dk\,
  |G(k)|^2 e^{-i\omega(k)t}\int_0^t
  dt^\prime\,e^{i\omega(k)t^\prime}\psi_a(t^\prime)-iA(t)
\end{equation}
where $A(t) \equiv \int_{-\pi}^\pi dk\, G(k) e^{-i\omega(k)t}\psi(k,0)$.
Next, we take the Laplace transform on both sides of the above equation by
defining $\tilde{\psi}_a(s) = \int_0^\infty dt\,e^{-st}\psi_a(t)$.  By also
using the fact that $\int_0^\infty dt\, e^{-st}\frac{\partial
\psi_a(t)}{\partial t} = s\tilde{\psi}_a(s) - \psi_a(0)$, and defining
$\tilde{A}(s)$ as the Laplace transfrom of $A(t)$, we arrive at
\begin{equation}
    s\tilde{\psi}_a(s) - \psi_a(0) = -i\omega_a \tilde{\psi}_a(s) -i
    \tilde{A}(s)-\int_{-\pi}^\pi dk\,|G(k)|^2\int_0^\infty dt\,
    e^{-(s+i\omega(k))t} F(t)
\end{equation}
where $F(t) \equiv \int_0^t dt^\prime
e^{i\omega(k)t^\prime}\psi_a(t^\prime)$.  Realising that $\frac{d}{dt}F(t) =
e^{i\omega(k)t}\psi_a(t)$, we perform integration by parts on $\int_0^\infty
dt\, e^{-(s+i\omega(k))t} F(t)$ and after some algebra finally arrive at
\begin{equation}
    \left(s+i\omega_a + i\Sigma(s)\right)\tilde{\psi}_a(s) = -i\tilde{A}(s) + \psi_a(0)
\end{equation}
where $\Sigma(s)$ as per defined in \eqref{eqn: Sigma(s) equation}.  By
inserting into the above equation the initial conditions corresponding to the
case of giant atom decay, $\psi_a(0) = 1$ and $\psi(k,0) = 0$ for all $k$, we
finally arrive at
\begin{equation}
    \tilde{\psi}_a(s) = \frac{1}{s + i\omega_a + i\Sigma(s)}
\end{equation}
which we can invert via the Bromwich integral to give us $\psi_a(t)$ in the following manner,
\begin{align*}
  \psi_a(t) &= \frac{1}{2\pi i} \int_{\lambda -i\infty}^{\lambda + i \infty}
  \frac{e^{st}}{s+i\omega_a + i\Sigma(s)} \nonumber \\
  &= \sum_{\text{All residues}} 
  \frac{e^{st}}{s+i\omega_a + i\Sigma(s)}
\end{align*}
where to go from the first line to the second line, we pick $\lambda$
sufficiently large so that all the poles of the integrand lie on the left of
the line $\lambda + it$, $t\in (-\infty,\infty)$.  The second line is
\eqref{eqn: decay dynamics} in the main text.  We note that in going from the
first line to the second line, we have ignored the contributions from the
integration over any paths induced by possible branch cuts.  This is because
we will mainly use the above equation to study the behavior of BICs, which
are poles on the imaginary axis and hence lead to long-term, non-decaying
behavior of $\psi_a(t)$.  The integration over any paths induced by possible
branch cuts leads to transient decay behavior which we are not interested in.

\section{Detailed calculations for section~\ref{sec: oscillating bound states} in the main text}
\label{Appendix: oscillating BIC calculations}
\subsection{Derivation of \eqref{eqn: sigma(s) for M legs oscillating BS}}
First of all, we note the important integral result 
\begin{align}
  \label{eqn: In(s) integral result}
  I_n(s) &= \int_{-\pi}^{\pi} dk\,\frac{e^{ikn}}{is - \cos(k)} \nonumber \\ 
  &= \frac{-2\pi i}{\sqrt{s^2+1}} \left(is \mp i \sqrt{s^2+1}\right)^{|n|}
\end{align}
where we have the minus sign when $\text{Re}(s)>0$ and the positive sign when
$\text{Re}(s)<0$.  The result can be derived by making the subtitution
$z=e^{ik}$ and thereafter computing the resultant complex integral
\begin{align}
  I_n(s) &= \oint_C dz \,\frac{z^n}{is - (1/2)(z+z^{-1})} \frac{1}{iz} \nonumber \\ 
  &=2i \oint_C dz\, \frac{z^n}{z^2 - 2isz +1} \nonumber \\
  &=2i \oint_C dz\, \frac{z^n}{(z-z_1)(z-z_2)}
\end{align}
where the contour $C$ is the unit circle in the complex plane and $z_1 =
i(s+\sqrt{s^2+1})$, $z_2 = i(s-\sqrt{s^2+1})$.  The above integral can then
be easily evaluated by the Cauchy Residue Theorem.  Now, we can show that
$\text{Re}(s)>0$ implies that $|z_1|>1$, $|z_2|<1$ which means that $z_2$ is
a simple pole in $C$.  On the other hand, $\text{Re}(s)<0$ implies that
$|z_1|<1$, $|z_2|>1$ which means that $z_1$ is a simple pole in $C$.
Moreover, if $n <0$ then we also have a $n$th order pole at z=0 in $C$.
Putting all of these together with the fact that $z_1 z_2 = 1$ which means
that $z_1^{-|n|} = z_2^{|n|}$, we can arrive at \eqref{eqn: In(s) integral
result} after some algebra.

Thereafter, using the above result we can easily derive the following
integral, with $n$ being a non-negative integer
\begin{align}
  \label{eqn: important cosine integral equation}
  \int_{-\pi}^{\pi} dk \frac{\cos(nk)}{is-2J\cos(k)} =  \frac{-2\pi i}{\sqrt{s^2+4J^2}} \left(\frac{is \mp i\sqrt{s^2 + 4J^2}}{2J}\right)^n
\end{align}
where again we have the minus sign when $\text{Re}(s)>0$ and the positive
sign when $\text{Re}(s) < 0$. Next, we can derive
\begin{align}
  |G(k)|^2 &= \frac{\rho^2}{2\pi} \sum_{j=0}^{M-1}\sum_{l=0}^{M-1}e^{i k n_0
  (j-l)} \nonumber \\
  &= \frac{\rho^2}{2\pi}\left(M + 2\sum_{r=1}^{M-1}(M-r)\cos(k n_0 r)\right)
\end{align}
where in going from the first line to the next, we realise that there are $M$
terms of $e^{i k (0) n_0}$, $(M-1)$ terms of $e^{ik(1)n_0}$ and its complex
conjugate, $(M-2)$ terms of $e^{ik(2)n_0}$ and its complex conjugate, and so
on, until at last we have $1$ term of $e^{ik(M-1)n_0}$ and its complex
conjugate.  With the form of $|G(k)|^2$ above and \eqref{eqn: important
cosine integral equation}, we can get \eqref{eqn: sigma(s) for M legs
oscillating BS} from \eqref{eqn: Sigma(s) equation}.

\subsection{Derivation of oscillating BIC conditions}
Since we are working with $\omega_a = 0$, we want to look for two BICs as
near the band centre as possible with energies $\pm \Omega_{\text{BIC}}$,
where $\Omega_{\text{BIC}}$ is a positive value to be determined.  Hence, we
want to minimise the quantity
\begin{equation}
  \label{eqn: quantity to be minimised BIC omega condition derivation}
  \left|\frac{\Omega}{2J}\right| = \cos\left(\frac{2\pi}{n_0}\left(m \pm
  \frac{1}{M}\right)\right)
\end{equation}
over all integer values of $m$.  Note that in the above expression, we have
already substituted \eqref{eqn: conditions for k for M legs oscillating BS}
into the dispersion relation $\omega(k) = 2J\cos(k)$.  One way to perform
this minimization is to first solve $m$ in the equation $|\Omega/2J|=0$ and
then rounding the value of $m$ to the closest integer.  When we solve
$|\Omega/2J|=0$, we obtain two cases
\begin{subequations}
  \begin{align}
    \frac{2\pi}{n_0}\left(m \pm \frac{1}{m}\right) = \frac{\pi}{2} \implies m
    = \frac{n_0}{4} \mp \frac{1}{m} \\
    \frac{2\pi}{n_0}\left(m \pm \frac{1}{m}\right) = \frac{3\pi}{2} \implies
    m = \frac{3n_0}{4} \mp \frac{1}{m}.
  \end{align}
\end{subequations}
At this juncture, before we round the value of $m$ obtained above to the
closest integer, we note that there are two cases we need to consider: either
$n_0 = 2(2l)$ or $n_0 = 2(2l+1)$ where $l \in \mathbb{Z}$.  In the former
case where $n_0 = 4l$, the closest integer value of $m$ would be $m = n_0/4$ or $m = 3n_0/4$,
which would give us the BIC frequencies in \eqref{eqn: BIC energies M legs
n0=4l} when we substitute those values of $m$ into \eqref{eqn: quantity to be
minimised BIC omega condition derivation}. Thereafter, we can obtain $\rho_0$ by substituting those frequencies into \eqref{eqn: delta(Omega) for M legs oscillating BS} to obtain \eqref{eqn: BIC rho0 M legs n0=4l}.

In the latter case where $n_0 = 4l+2$, the closest integer values of $m$
would be $m = n_0/4 \pm 1/2$ or $m = 3n_0/4 \pm 1/2$, corresponding to two
possible BIC frequencies,
\begin{equation*}
  \Omega_{\text{BIC}} = \sin\left(\frac{\pi}{n_0}\left(1\pm \frac{2}{M}\right)\right).
\end{equation*}
However, when we substitute $\Omega_{\text{BIC}} =\sin((\pi/n_0)(1-2/M))$
into \eqref{eqn: delta(Omega) for M legs oscillating BS}, for the case of
$\omega_a=0$, we end up with $\rho_0^2<0$, which means that we are only left
with $\Omega_{\text{BIC}} =\sin((\pi/n_0)(1+2/M))$ for the $n_0 = 4l+2$ case.
However, since this is further from the band center as compared to the $n_0 =
4l$ case, the resultant bound states would have lower emitter probability. Hence, we can conclude that $n_0 = 4l$ would give optimal oscillating BICs.

As an example, for the case where $M=3$ and $n_0 = 4l+2, n_0>2$, we have the BIC
frequency $|\Omega| = 2 J \sin \left( \frac{5\pi}{3n_0} \right)$ which gives us 
\begin{equation}
    \rho_0^2 = \frac{2J^2}{\sqrt{3}} \sin \left( \frac{10\pi}{3 n_0} \right)
\end{equation}
Using \eqref{eqn: emitter contribution BIC}, we find that the emitter
probability
\begin{equation}
    |\phi_a^{\text{BIC}}|^2 \to \frac{9}{9+20\sqrt{3}\pi} \approx 0.0764
\end{equation}
as $n_0 \to \infty$, which is considerably smaller than the optimal value
$\approx 0.171$ when $n_0$ is an integer multiple of $4$ for $M=3$.

\subsection{Derivation of \eqref{eqn: asymptotic BOC energies} and \eqref{eqn: asymptotic BOC probability}}
To derive \eqref{eqn: asymptotic BOC energies}, the method is to solve
\eqref{eqn: sigma(s) for M legs oscillating BS} using both \eqref{eqn: big
term in sigma(s) expression} and \eqref{eqn: detuning in terms of Sigma(s)}
for the case where $\omega_a =0$ and $|\Omega| > 2J$.  The $|\Omega| > 2J$
condition is because we are trying to solve for $\Omega$ outside the energy
band.  We note that there are two cases for us to consider, namely
$\text{Re}(s) > 0$, corresponding to $s = -i \Omega + 0^+$ and $\text{Re}(s)
< 0$, corresponding to $s = -i\Omega - 0^+$.  As we will see, the
$\text{Re}(s) >0$ case gives us the BOC energy for $\Omega < -2J$, and the
$\text{Re}(s) < 0$ case gives use the BOC energy for $\Omega > 2J$.  First,
we consider the $\text{Re}(s) > 0$ case.  In this case, we have the equation
\begin{equation}
  \left(\frac{\Omega}{2J}\right)^2 = -\frac{1}{4}\left(\frac{\rho_0}{J}\right)^2 \frac{1}{\sqrt{\left(\frac{\Omega}{2J}\right)^2-1}}\left(M + 2\sum_{r=1}^{M-1} (M-r) \left(\frac{\Omega}{2J} + \sqrt{\left(\frac{\Omega}{2J}\right)^2-1}\right)^{n_0 r}\right)
\end{equation}
where we have used $\sqrt{4J^2 - \Omega^2} = i\sqrt{\Omega^2 - 4J^2}$, since
$|\Omega|>2J$. Now, we substitute $\Omega^\prime = \Omega/2J$ as well as \eqref{eqn: BIC rho0 M legs n0=4l} for $\rho_0/J$ to get
\begin{equation}
  \Omega^\prime \sqrt{{\Omega^\prime}^2 - 1} = -\frac{1}{4}\frac{2}{M}\tan\left(\frac{\pi}{M}\right)\sin\left(\frac{4\pi}{M n_0}\right) \left(M + 2\sum_{r=1}^{M-1} (M-r)\left(\Omega^\prime + \sqrt{{\Omega^\prime}^2 - 1}\right)^{n_0 r}\right)
\end{equation}
In the $n_0 \to \infty$ limit, for the above equation to have a solution, we
must have $\Omega^\prime + \sqrt{{\Omega^\prime}^2 - 1} < 1$, which means $\Omega < -2J$.  Hence, taking the $n_0 \to \infty$ limit, we have 
\begin{equation}
  \Omega^\prime \sqrt{{\Omega^\prime}^2 - 1} = -\frac{2}{M}\tan\left(\frac{\pi}{M}\right)\left(\frac{\pi}{Mn_0}\right)M
\end{equation}
Solving for $\Omega^\prime$ in the above equation subject to the condition that $\Omega < -2J$, we have 
\begin{align}
  \Omega^\prime &= -\sqrt{\frac{1}{2} \sqrt{\frac{16 \pi ^2 \tan
  ^2\left(\frac{\pi }{M}\right)}{M^2 \text{n0}^2}+1}+\frac{1}{2}} \nonumber \\ 
  &\approx -1 - \frac{2 \pi^2 \tan\left(\frac{\pi}{M}\right)^2}{M^2 n_0^2}.
\end{align}
The last line in the above equation is \eqref{eqn: asymptotic BOC energies}
for the case where $\Omega < -2J$.  The derivation for the $\Omega > 2J$ case
can be done by choosing $\text{Re}(s) < 0$ and following the same steps above
to arrive at
\begin{align}
  \Omega^\prime &= \sqrt{\frac{1}{2} \sqrt{\frac{16 \pi ^2 \tan
  ^2\left(\frac{\pi }{M}\right)}{M^2 \text{n0}^2}+1}+\frac{1}{2}} \nonumber \\ 
  &\approx 1 + \frac{2 \pi^2 \tan\left(\frac{\pi}{M}\right)^2}{M^2 n_0^2}
\end{align}
where the last line in the above equation is \eqref{eqn: asymptotic BOC
energies} for the case where $\Omega > 2J$.  To derive \eqref{eqn: asymptotic
BOC probability}, the method is to use \eqref{eqn: emitter contribution BIC}
together with \eqref{eqn: sigma(s) for M legs oscillating BS} and \eqref{eqn:
detuning in terms of Sigma(s)}.  We will also use \eqref{eqn: asymptotic BOC
energies} derived above, and also \eqref{eqn: BIC rho0 M legs n0=4l} for
$\rho_0/J$.  Thereafter, doing an asymptotic expansion about $\frac{1}{n_0}
\to 0$ and keeping the lowest order, we obtain \eqref{eqn: asymptotic BOC
probability}.

\subsection{Derivation of \eqref{eqn: M=3 bic probability dist}}
Firstly, we write our BICs $\ket{\pm}$ as 
\begin{align}
  \ket{\pm} &= \phi_a^{(\pm)} \ket{1_a} + \int_{-\pi}^\pi dk \, e(k)^{\pm}
  \ket{1_k} \nonumber \\
  &= \phi_a^{(\pm)} \ket{1_a} + \phi_a^{(\pm)} \int_{-\pi}^\pi dk \, \frac{G^*(k)}{\Omega_{\pm} - 2J \cos(k)} \ket{1_k}
\end{align}
where $\phi_a^{(\pm)} = \braket{1_a}{\pm}$ and $e(k)^{\pm} =
\braket{1_k}{\pm}$, and $\Omega_{\pm}$ is the energy of the $\ket{\pm}$ state
respectively.  In going from the first line to the second line, we used
\eqref{eqn: one excitation subspace energy eigval eqn simultaneous eqn}.  In
the second line above, we see that we can clearly factor out a global phase
corresponding to the phase of $\braket{1_a}{\pm}$.  Hence, without loss of
generality, we can write $\phi_a \equiv \phi_a^{(\pm)} = \braket{1_a}{\pm}$.
Then, we have
\begin{align}
  \ket{\pm} 
  &= \phi_a \ket{1_a} + \phi_a \int_{-\pi}^\pi dk \,
  \frac{G^*(k)}{\Omega_{\pm} - 2J \cos(k)} \ket{1_k} \nonumber\\
  &= \phi_a \ket{1_a} + \phi_a \int_{-\pi}^\pi dk \,
  \frac{G^*(k)}{\Omega_{\pm} - 2J \cos(k)} c(k)^\dagger\ket{0} \nonumber\\
  &= \phi_a \ket{1_a} + \phi_a \int_{-\pi}^\pi dk \,
  \frac{G^*(k)}{\Omega_{\pm} - 2J \cos(k)} \sum_n e^{i k
  n}\frac{1}{\sqrt{2\pi}}b_n^\dagger\ket{0} \nonumber\\
  &= \phi_a \ket{1_a} + \phi_a \int_{-\pi}^\pi dk \,
  \frac{(\rho_0/2\sqrt{\pi})\sum_{l=0}^{M-1} e^{-i k n_l}}{\Omega_{\pm} - 2J \cos(k)} \sum_n e^{i k
  n}\frac{1}{\sqrt{2\pi}}b_n^\dagger\ket{0} \nonumber\\
  &= \phi_a \ket{1_a} +  \sum_n
  \underbrace{\frac{\rho_0\phi_a}{2\pi}\sum_{l=0}^{M-1}\int_{-\pi}^\pi dk \, \frac{ e^{i k
  (n-n_l)}}{\Omega_{\pm} - 2J \cos(k)}}_{\equiv \phi_{n, \pm}} b_n^\dagger\ket{0}
\end{align}
Thereafter, to evaluate $\phi_{n,\pm}$, we note that 
\begin{align}
  \phi_{n,\pm} &= \frac{\rho_0\phi_a}{2\pi}\sum_{l=0}^{M-1}\int_{-\pi}^\pi dk
  \, \frac{ e^{i k (n-n_l)}}{\Omega_{\pm} - 2J \cos(k)} \nonumber \\
  &=  \frac{\rho_0\phi_a}{2\pi}\sum_{l=0}^{M-1} I_{n-n_l}(s=-i\Omega_\pm+0^+)
\end{align}
where $I_{n-n_l}(s)$ is given by \eqref{eqn: In(s) integral result}.

\section{Proof that there is no oscillating BIC with
\texorpdfstring{$M=2$}{M=2}}
\label{Appendix: oscillating BIC at n=2 CMI}
For $n=2$, we have 
\begin{equation}
  |G(k)|^2 = \frac{\rho_0^2}{\pi}\left(1 + \cos(kn_0)\right)
\end{equation}
which gives us 
\begin{equation}
  \Sigma(s) = \frac{-2i
  \rho_0^2}{\sqrt{s^2+4J}}\left(1+\left(\frac{-i\sqrt{s^2+4J^2}+is}{2J}\right)^{n_0}\right).
\end{equation}
Thus, we have 
\begin{align}
  \Sigma(s=-i\Omega) &= \frac{-2i
  \rho_0^2}{\sqrt{4J-\Omega^2}}\left(1+\left(\frac{-i\sqrt{4J^2-\Omega^2}+\Omega}{2J}\right)^{n_0}\right) 
  \nonumber \\
  &= \frac{-2i \rho_0^2}{\sqrt{4J-\Omega^2}}\left(1+\cos(n_0
  \theta)+i\sin(n_0\theta)\right)
\end{align}
where $\theta = \arctan\left(\frac{-\sqrt{4J^2-\Omega^2}}{\Omega}\right)$ and
$-2J<\Omega < 2J$.
Using $\Delta(\Omega) = \text{Re}(\Sigma(s=-i\Omega +  0^+))$, we have
\begin{equation}
    \label{eqn: oscillating bs n=2 detuning}
    \Omega - \omega_a = \frac{2\rho_0^2}{\sqrt{4J-\Omega^2}}\sin(n_0 \theta)
\end{equation}
Furthermore, enforcing $|G(k)|^2=0$ gives us 
\begin{equation}
  k = (2l+1)\frac{\pi}{n_0}
\end{equation}
where $l$ is an integer. Hence, we have 
\begin{equation}
    \Omega = 2J\cos\left(\frac{\pi}{n_0}(2l+1)\right)
\end{equation}
which we can substitute into $\theta = \arctan\left(\frac{-\sqrt{4J^2-\Omega^2}}{\Omega}\right)$ to get 
\begin{equation}
    \theta = -\frac{\pi}{n_0}(2l+1)
\end{equation}
Subtituting the above expression into \eqref{eqn: oscillating bs n=2 detuning}, we have 
\begin{equation}
  \Omega = \omega_a.
\end{equation}
Hence, when we have $K=2$ legs in our giant atom, there is only one BIC at
the frequency $\Omega = \omega_a$, provided that
\begin{equation}
  \omega_a = \Omega = 2J\cos\left(\frac{\pi}{n_0}(2l+1)\right).
\end{equation}
Otherwise, there is no BIC for $K=2$ legs.  Moreover, since we only have one
BIC, it is impossible to get an oscillating BIC, which requires at least two
BICs at two different frequencies in the band.

\section{Comparison of the oscillating BIC conditions with continuous waveguide}

For $M$ coupling points, and setting $\omega_a = 0$, the oscillating BIC found for a continuous (linearized) waveguide in Ref.~\cite{guo_oscillating_2020} are formed from the superposition of two BICs with energies
\begin{equation}
    \Omega_{c} = \pm \frac{1}{2} M \gamma \cot\left( \frac{n\pi}{M} \right), \quad n=1,2,\ldots,\left\lfloor M/2 \right\rfloor
\end{equation}
where $\gamma$ is the giant atom decay rate into the waveguide. In our case, $\gamma$ corresponds to $\rho_0^2/J$. From~\eqref{eqn: BIC energies M legs n0=4l} and~\eqref{eqn: BIC rho0 M legs n0=4l}, we have, for our case,
\begin{equation}
    \Omega_{\text{BIC}} = \pm 2J \sin\left( \frac{2\pi}{Mn_0} \right) = \pm 2 \gamma \frac{J^2}{\rho_0^2} \sin\left( \frac{2\pi}{Mn_0} \right) = \pm \frac{1}{2} M\gamma \cot\left( \frac{\pi}{M} \right) \sec\left( \frac{2\pi}{Mn_0} \right)
\end{equation}
which looks similar to the continuous-waveguide result for $n=1$. In fact, in the regime where $(M n_0)$ is large such that $\sec(2\pi/(Mn_0))\approx 1$, 
the oscillating BIC energies are approximately the same as $\Omega_c$.

\end{document}